\documentclass[aps,twocolumn,pre,superscriptaddress]{revtex4-1}
\usepackage{amssymb}
\usepackage{amsmath}
\usepackage{color}
\usepackage{sidecap}
\usepackage{graphicx}

\usepackage{xcolor,hyperref}
\hypersetup{
   colorlinks,
   linkcolor={blue!50!black},%{red!80!black},
   citecolor={blue!50!black},
   urlcolor={blue!80!black}
}
\DeclareMathAlphabet{\mathitbf}{OML}{cmm}{b}{it}
\newcommand{\dbar}{{\,\mathchar'26\mkern-12mu d}}

\newcommand{\xv}{\mathitbf x}

\newcommand{\sFrac}[2]{{\textstyle\frac{#1}{#2}}}

\newcommand{\calBold}[1]{\mbox{\boldmath${\cal #1}$}}
\newcommand{\mathBold}[1]{\mbox{\boldmath$#1$}}

\begin{document}

\title{Universality of the nonphononic vibrational spectrum\\ across different classes of computer glasses}

\author{David Richard}
\thanks{contributed equally}
\affiliation{Institute for Theoretical Physics, University of Amsterdam, Science Park 904, 1098 XH Amsterdam, The Netherlands}
\author{Karina Gonz\'alez-L\'opez}
\thanks{contributed equally}
\affiliation{Institute for Theoretical Physics, University of Amsterdam, Science Park 904, 1098 XH Amsterdam, The Netherlands}
\author{Geert Kapteijns}
\affiliation{Institute for Theoretical Physics, University of Amsterdam, Science Park 904, 1098 XH Amsterdam, The Netherlands}
\author{Robert Pater}
\affiliation{Institute for Theoretical Physics, University of Amsterdam, Science Park 904, 1098 XH Amsterdam, The Netherlands}
\author{Talya Vaknin}
\affiliation{Chemical and Biological Physics Department, Weizmann Institute of Science, Rehovot 7610001, Israel}
\author{Eran Bouchbinder}
\affiliation{Chemical and Biological Physics Department, Weizmann Institute of Science, Rehovot 7610001, Israel}
\author{Edan Lerner}
\affiliation{Institute for Theoretical Physics, University of Amsterdam, Science Park 904, 1098 XH Amsterdam, The Netherlands}

\begin{abstract}
It has been recently established that the low-frequency spectrum of simple computer glass models is populated by soft, quasilocalized nonphononic vibrational modes whose frequencies $\omega$ follow a gapless, universal distribution ${\cal D}(\omega)\!\sim\!\omega^4$. While this universal nonphononic spectrum has been shown to be robust to varying the glass history and spatial dimension, it has so far only been observed in simple computer glasses featuring radially-symmetric, pairwise interaction potentials. Consequently, the relevance of the universality of nonphononic spectra seen in simple computer glasses to realistic laboratory glasses remains unclear. Here we demonstrate the emergence of the universal $\omega^4$ nonphononic spectrum in a broad variety of realistic computer glass models, ranging from tetrahedral network glasses with three-body interactions, through molecular glasses and glassy polymers, to bulk metallic glasses (BMGs). Taken together with previous observations, our results indicate that the low-frequency nonphononic vibrational spectrum of any glassy solid quenched from a melt features the universal $\omega^4$ law, independently of the nature of its microscopic interactions.
\end{abstract}

\maketitle

\section{introduction}

It is common in condensed matter physics that dynamic and thermodynamic phenomena are controlled by low-energy excitations \cite{kittel2005introduction,tsvelik2003quantum}. For example,
in crystalline solids, phonon-phonon interactions control wave attenuation rates and heat transport \cite{ketterson2016physics}; dislocations (i.e.~low-energy topological defects) mediate plastic deformation rates upon external mechanical loading \cite{suzuki2013dislocation}; the specific heat grows as the third power of temperature due to the $\sim\!\omega^2$ Debye distribution of phonon frequencies. The same principle is also seen to hold in glassy solids, in which soft two-level systems, and their interactions with phonons, are believed to control thermodynamic and transport properties below 10K \cite{Zeller_and_Pohl_prb_1971,Anderson,Phillips}, and low-energy, quasilocalized excitations --- often referred to as shear transformation zones \cite{falk_langer_stz} --- govern elasto-plastic responses \cite{plastic_modes_prerc}. Consequently, the complete understanding of the statistical mechanics of soft excitations in solids, and in particular in glasses, is of key importance.

\begin{figure}[ht!]
  \includegraphics[scale=1]{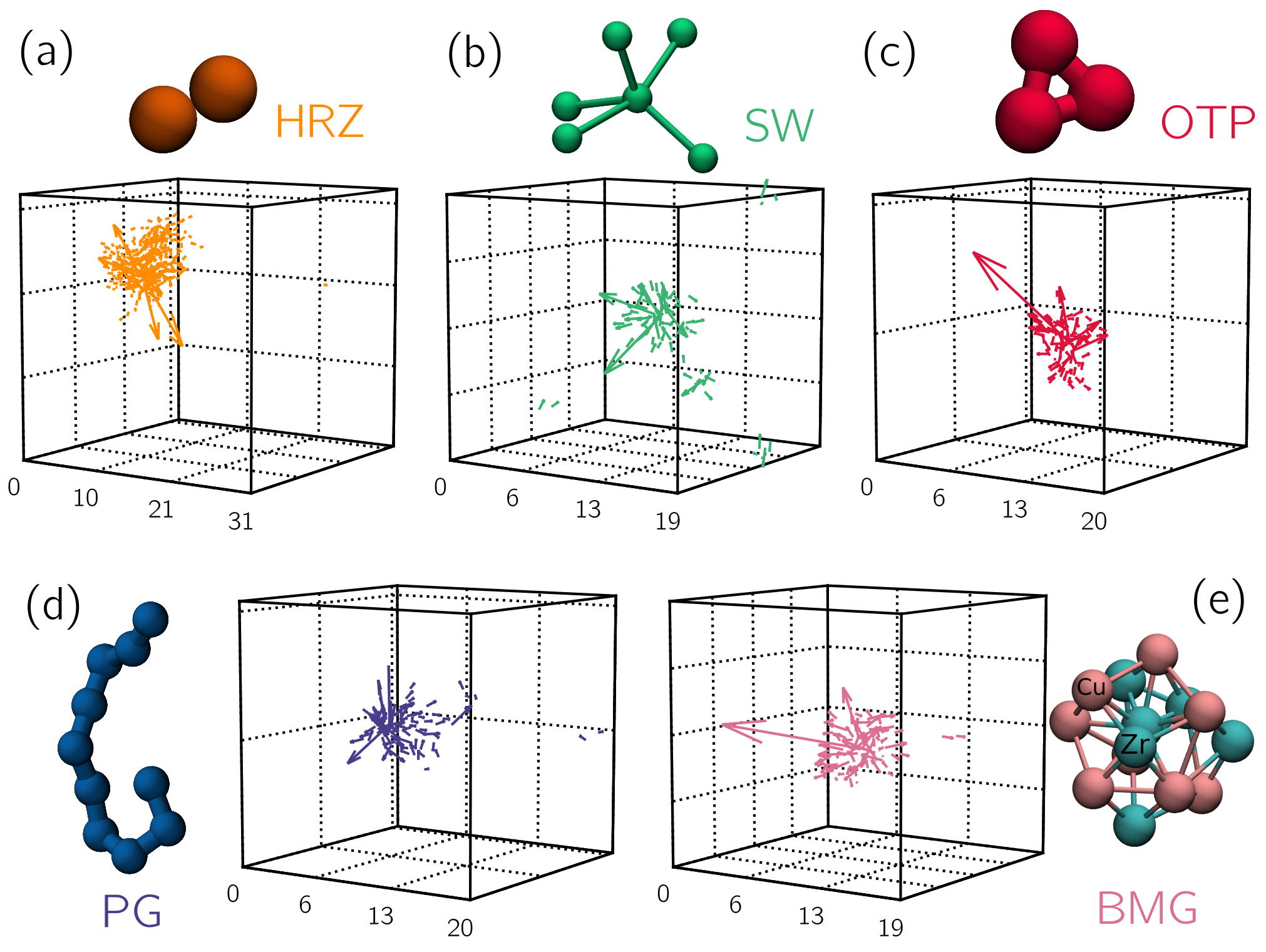}
  \caption{\footnotesize In this work we study five realistic glass forming models, each representing a different class of disordered solids, as illustrated by the cartoons. Visualizations of quasilocalized modes found in the employed models of (a) an elastic-spheres glass, (b) a network glass, (c) a molecular glass, (d) a polymer glass, and (e) a bulk metallic glass. For visualization purposes, only the largest 1\% of components are shown.}
  \label{fig:1}
\end{figure}

Indeed, much attention has been devoted in the past few decades to understanding the low-frequency spectra of glassy solids \cite{soft_potential_model_1991,Gurevich2003,Gurevich2007,barrat_3d,Schirmacher_prl_2007,experiments_1620K_vSiO2,mw_EM_epl,sokolov_boson_peak_scale,Monaco_prl_2011,eric_boson_peak_emt, eric_hard_spheres_emt, silvio, modes_prl_2016, modes_prl_2018, cge_paper, pinching_pnas, ikeda_pnas, LB_modes_2019, atsushi_core_size_pre}. It is now well-accepted that soft, quasilocalized modes dwell at vanishing frequencies $\omega\!\to\!0$ in simple computer glasses. These nonphononic excitations were shown to universally feature a disordered core of linear size of about 10 particle diameters \cite{pinching_pnas} (see examples in Fig.~\ref{fig:1}), decorated with algebraically-decaying (mostly-affine) displacement fields of magnitude $\sim\! r^{-(\dbar-1)}$ at distance $r$ away from the core, in $\dbar$ spatial dimensions. The frequencies associated with these excitations were shown to follow a universal distribution ${\cal D}(\omega)\!\sim\!\omega^4$  \cite{modes_prl_2016}, independent of spatial dimension \cite{modes_prl_2018} or preparation protocol \cite{cge_paper,LB_modes_2019,pinching_pnas}. While these numerical observations are supported by various theoretical frameworks \cite{soft_potential_model_1991,Gurevich2003}, their relevance for laboratory glasses has not been well-established; to the best of our knowledge, all computational investigations of the asymptotic functional form of low-frequency nonphononic spectra to date (with the exception of \cite{bonfanti2020universal} put forward in parallel to this work) employed simple computer glass models, in which particles interact via radially-symmetric, pairwise potentials.

In this work we create {\em in silico} ensembles of polymeric glasses, tetrahedal network glasses, elastic sphere glasses, molecular glasses and bulk metallic glasses (BMGs), which are considerably more realistic representatives of laboratory glasses compared to the simple computer-glass models investigated previously, in order to test whether the universal nonphononic spectrum observed in simple computer glasses remains relevant to laboratory glasses as well. Our main finding is that these realistic glass-forming models also feature the universal ${\cal D}(\omega)\!\sim\!\omega^4$ nonphononic vibrational density of states (vDOS), as seen in simple computer glasses. We thus extend the degree of universality of the $\omega^4$ law, and lend substantial support to the assertion that any glass formed by quenching a melt --- and, in particular, laboratory glasses --- would feature the gapless $\omega^4$ nonphononic vDOS.

\section{Computer glass models}

We employ five computer glass models, each representing a different \emph{class} of glassy solids. Here we briefly review the employed models, keeping a complete description for Appendix~\ref{models_appendix}. Throughout this work we express frequencies in terms of $c_s/a_0$, where $c_s$ is the shear wave speed, and $a_0$ is the typical interparticle distance, both are precisely defined in Appendix~\ref{models_appendix}.

The employed models are as follows:

(1) An elastic-spheres glass model in which spherical particles interact via the linear-elastic Hertz contact law \cite{hertz2006beruhrung}. At low confining pressures, this model undergoes an unjamming transition \cite{ohern2003,liu_review,van_hecke_review}. We refer to this model as HRZ.

(2) The Stillinger-Weber network glass model \cite{Stillinger_Weber}, which employs a three-body term in the potential energy that favors tetrahedral local structures. In some range of its parameters, this model mimics the behavior of amorphous silica \cite{sw_vary_lambda}. We refer to this model as SW.

(3) A triatomic molecular glass model inspired by glass-forming models of orthoterphenyl \cite{otp_paper,LEWIS1993295}, referred to in what follows as OTP.

(4) A polymer-glass model of soft beads connected by FENE nonlinear springs \cite{Starr2002}, referred to in what follows as PG. Monomers between different polymers interact with a Lennard-Jones-like potential \cite{fsp}.

(5) A binary bulk metallic glass (BMG) alloy composed of Copper (Cu) and Zirconium (Zr) atoms according to Cu$_{\rm 46}$Zr$_{\rm 54}$~\cite{cheng2009atomic,cheng2011atomic}. The interactions are calculated using the Embedded-Atom Method (EAM), which gives rise to a spherically-symmetric, many-body potential.

Detailed descriptions about how ensembles of glassy samples were created for each computer glass model are provided in Appendix~\ref{models_appendix}. Briefly described, we generate uncorrelated equilibrium configurations at temperatures much larger than $T_g$, and perform an energy minimization on those configurations to obtain zero-temperature glassy solids. 

%Due to the different masses of the particles the modes of the system are given by diagonalizing the dynamical matrix
% \begin{equation}
% \calBold{D}_{ij} = \frac{\calBold{M}_{ij}}{\sqrt{m_im_j}}\ ,
% \end{equation}

% $(\cdot)$ denotes a contraction over spatial components, and ${\cal T}$ is a diagonal matrix of the particles' masses
%  \[
%  {\cal T}  =
%   \begin{bmatrix}
%     m_{1} & & \\
%     & \ddots & \\
%     & & m_{N}
%   \end{bmatrix}\,.
% \]

For each generated glassy sample, we perform a normal mode analysis, which follows from a generalized eigenvalue problem: eigenvectors $\mathBold{\psi}$ and eigenfrequencies $\omega$ satisfy the equation 
\begin{equation}\label{foo00}
\sum_j\calBold{M}_{ij}\cdot\mathBold{\psi}_j = m_i \omega^2 \mathBold{\psi}_i\,.
\end{equation}
Here $m_i$ denotes the mass of the $i^{\mbox{\tiny th}}$ particle, the Hessian matrix reads $\calBold{M}_{ij}\!\equiv\!\frac{\partial^2 U}{\partial \xv_i \partial \xv_j}$, where $U$ denotes the potential energy and $\xv_i$ is the $\dbar$-dimensional coordinate vector of the $i^{\mbox{\tiny th}}$ particle, and $\mathBold{\psi}_j$ is the $\dbar$-dimensional Cartesian displacement vector of the $j^{\mbox{\tiny th}}$ particle. Note that no summation over $i$ is implied on the right-hand-side of Eq.~(\ref{foo00}). To obtain the eigenvectors $\mathBold{\psi}$ and eigenfrequencies $\omega$, we solve the auxiliary eigenvalue problem
\begin{equation}
\sum_j\frac{\calBold{M}_{ij}}{\sqrt{m_im_j}}\cdot\calBold{\phi}_j = \omega^2\calBold{\phi}_i\,.
\end{equation}
where $\mathBold{\psi}_i\!=\!\calBold{\phi}_i/\sqrt{m_i}$. Details about the calculation of $\calBold{M}$ for the SW and BMG systems are provided at length in Appendices~\ref{SW_appendix} and \ref{BMG_appendix}. The system and ensemble sizes in our simulations were selected such that the lowest-frequency modes appear below the first phononic band, as explained in detail in~\cite{modes_prl_2016,finite_size_modes}.

\begin{figure*}[t!]
  \includegraphics[scale=1]{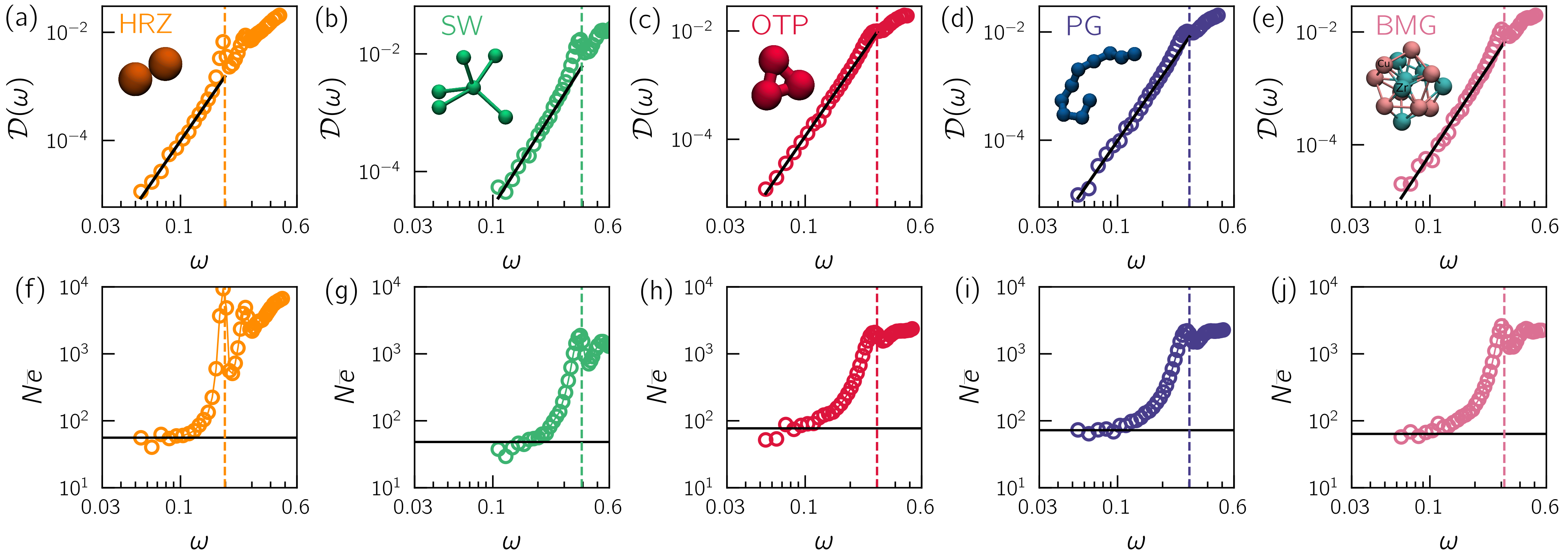}
  \caption{ \footnotesize Low-frequency vibrational modes' spectra and localization properties, measured in realistic computer glass models. We show ${\cal D}(\omega)$ vs.~frequency $\omega$ for a linear-elastic-spheres glass (HRZ, panel (a)), a network glass (SW, panel (b)), a molecular glass (OTP, panel(c)), a polymer glass (PG, panel (d)), and a bulk metallic glass (BMG, panel(e)). The solid lines indicate ${\cal D}(\omega)\!\sim\!\omega^4$. Panels (f)-(j) show the average participation ratio $\bar{e}$ (see definition in Eq.~(\ref{participation_definition})), scaled by the number of particles $N$, binned over and plotted against frequency, for the same models of panels (a)-(d) respectively. The vertical dashed lines mark the first phonon band frequency $2\pi c_s/L$. The horizontal lines represent estimations $Ne_0$ of the low-frequency plateau, which capture the core size of soft quasilocalized modes, see values reported in Fig.~\ref{fig:3}a and text for further discussions.}
  \label{fig:2}
\end{figure*}

\section{results}

Our key result is displayed in Fig.~\ref{fig:2}a-d, where we show the low-frequency regime of the vDOS of all simulated computer glasses. All models feature the universal form ${\cal D}(\omega)\!\sim\!\omega^4$, despite the stark qualitative differences between the microscopic interaction laws that define each model. 

A quantitative comparison of the localization properties of quasilocalized modes between our various computer models is made possible by studying those modes' participation ratio
\begin{equation}\label{participation_definition}
e\equiv \frac{(\sum_i\mathBold{\psi}_i\cdot\mathBold{\psi}_i)^2}{N\sum_i (\mathBold{\psi}_i\cdot\mathBold{\psi}_i)^2}\,,
\end{equation}
where $\mathBold{\psi}_i$ denotes the $\dbar$-dimensional vector of a mode's Cartesian components pertaining to the $i^{\mbox{\tiny th}}$ particle. The participation ratio is expected to scale as $1/N$ for localized modes \cite{footnote}, and should be of order unity for extended modes (e.g.~phonons). The product $Ne$ is thus expected to reflect the core size of quasilocalized modes, expressed in terms of the characteristic volume occupied by a single particle.
\begin{figure}[ht!]
  \includegraphics[scale=1]{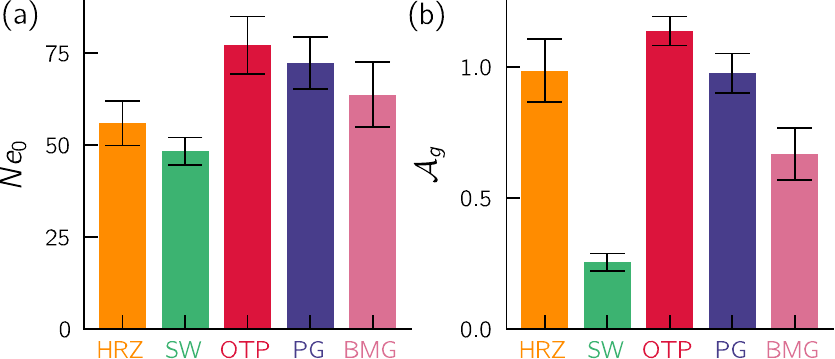}
  \caption{\footnotesize Dimensionless characterizers of the nonphononic vDOS, compared across different classes of glassy solids. (a) Low-frequency plateau $N e_0$ of the frequency-binned participation ratio $\bar{e}$ scaled by $N$, which represents the core size of soft, quasilocalized modes. (b) Dimensionless prefactors ${\cal A}_g\!\equiv\! A_g/(a_0/c_s)^5$ of the universal ${\cal D}(\omega)\!=\! A_g\omega^4$ nonphononic vDOS.}
  \label{fig:3}
\end{figure}

In Fig.~\ref{fig:2}e-h we show the mean participation ratio $\bar{e}$ of vibrational modes, scaled by system size $N$, binned over and plotted against frequency for all employed computer glasses. The first phonon band frequency $2\pi c_s/L$ is indicated by the vertical dashed lines, and features $Ne$ of order of a few thousands, consistent with the system sizes employed. Approaching zero frequency, we see that $N\bar{e}$ plateaus at a typical value $Ne_0$ on the order of a few tens, as marked by the horizontal dashed lines. The estimated values of the plateaus, $Ne_0$, are reported for all investigated computer glass models in Fig.~\ref{fig:3}a. Remarkably, the variation of $Ne_0$ across the different models is very small, of less than a factor of two with respect to each other.

Finally, we note that the prefactor $A_g$ of the nonphononic vDOS, namely ${\cal D}(\omega)\! =\! A_g\omega^4$, is an observable with dimensions of an inverse frequency to the fifth power. $A_g$ was discussed at length in \cite{cge_paper, pinching_pnas}, where it was argued to encompass information both about the number density of soft, quasilocalized modes, and about their characteristic stiffness. In those references it was shown that $A_g$ can be very sensitive to glass history, particularly for glasses that were deeply supercooled prior to their quench to the glass. Here we compare ${\cal A}_g\!\equiv\! A_g/(a_0/c_s)^5$ across our different computer glasses. The results are displayed in Fig.~\ref{fig:3}b; we find that ${\cal A}_g$ is of order unity in all models, with the exception of the SW network glass model that features ${\cal A}_g\!\approx\!0.2$.

We note that the quantities $Ne_0$ and ${\cal A}_g$ generally depend on glass history \cite{protocol_prerc,cge_paper,pinching_pnas}. However, these dependencies are most pronounced for glasses quenched from deeply supercooled liquids, and are generally weak for glasses quenched from high temperature liquid states.  Since in this work we indeed compare glasses quenched from high temperature liquid states, the history dependence of $Ne_0$ and ${\cal A}_g$ is expected to be weak, and therefore the comparison between them across different classes of glass-forming models is meaningful. We conclude that the energy landscapes of the computer glasses we investigate here share \emph{quantitative} similarities that extend beyond the universal scaling of their nonphononic vDOS.

\section{summary and outlook}

In this work we have shown that the low-frequency nonphononic spectra of realistic computer glass models --- including network glasses, polymer glasses, and molecular glasses --- feature the universal gapless $\omega^4$ law, as seen previously in simple computer glass models \cite{modes_prl_2016, ikeda_pnas, modes_prl_2018, cge_paper, LB_modes_2019, pinching_pnas}. We thus expand the degree of universality of the $\omega^4$ law to include several qualitatively different classes of realistic glass forming models, and reinforce its relevance to laboratory glasses. Finally, our results support the description of glasses' vibrational properties via mesoscale, coarse grained approaches that consider interacting oscillators and anharmonicities \cite{Gurevich2003,Gurevich2007}, in which the microscopic details play no role in determining the scaling with frequency of the nonphononic vDOS. We note that after the completion of this work, we became aware of the results obtained by Bonfanti et al.~\cite{bonfanti2020universal}, which support our conclusions. 

Our results underline the timeliness of formulating a first-principles theory that explains the observed universality of nonphononic spectra in glassy solids. Mean-field approaches that are based on a microscopic description \cite{eric_boson_peak_emt,eric_hard_spheres_emt,silvio} (rather than a coarse-grained one) predict that the nonphononic vDOS of glassy solids should scale as $\omega^2$, independent of spatial dimension. An important goal for future studies will be to consolidate the predictions of the mesoscopic \cite{Gurevich2003,Gurevich2007} and microscopic \cite{eric_boson_peak_emt,eric_hard_spheres_emt,silvio} theoretical approaches.

\acknowledgements
We wish to acknowledge inspiring discussions with Itamar Procaccia, Corrado Rainone, and Yuri Lubomirsky. D.~R.~acknowledges support of the Simons Foundation for the ``Cracking the Glass Problem Collaboration" Award No.~348126. K.~G.-L. acknowledges the computer resources provided by the Laboratorio Nacional del Sureste de M{\'e}xico, CONACYT member of the national laboratories network. E.~L.~acknowledges support from the NWO (Vidi grant no.~680-47-554/3259). E.~B.~acknowledges support from the Minerva Foundation with funding from the Federal German Ministry for Education and Research, the Ben May Center for Chemical Theory and Computation, and the Harold Perlman Family.

%% ---- bibliography ----

%\newpage

\appendix

\section{Computer glass models}
\label{models_appendix}

In this Appendix we provide detailed descriptions of the computer models employed in our work, and the protocol used to prepare our ensembles of glassy samples. For each model we specify its associated microscopic units; however, we reiterate that in the main text we express frequencies in terms of $c_s/a_0$ where $a_0\!\equiv\!(V/N)^{1/3}$ is a microscopic length, $c_s\!\equiv\!\sqrt{G/\rho}$ is the speed of shear waves, $G$ denotes the athermal shear modulus \cite{lutsko}, $\rho$ is the mass density, $V\!=\! L^3$ is the volume, and $N$ is the number of particles. For all models, the low-frequency spectra is extracted via a partial diagonalization using the ARPACK package~\cite{arpack}.

\subsection{Elastic spheres}
We employ a simple-yet-realistic model of soft, linear-elastic spheres interacting via the Hertzian interaction law \cite{hertz2006beruhrung}
\begin{equation}
\varphi_{\mbox{\tiny Hertz}}(r) = \sFrac{2\varepsilon}{5} \big( r - \sFrac{\sigma_i + \sigma_j}{2} \big)^{5/2}\Theta\big(\sFrac{\sigma_i + \sigma_j}{2} - r \big)\,,
\end{equation}
where $\sigma_i,\sigma_j$ denote the radii of the $i^{\mbox{\tiny th}}$ and $j^{\mbox{\tiny th}}$ particles, and $\Theta(x)$ is the Heaviside step function. We enclose $N\!=\!32000$ particles of equal mass $m$ in a box of volume $V\!=\! L^3$, and fix the number density at $N/V\!=\!0.9386$. We choose 50\% of particles to have $\sigma_i\!=\!0.5$ and the other 50\% to have $\sigma_i\!=\!0.7$. Length are expressed in terms of the diameter of the smaller species, and energies in terms of $\varepsilon$.

To make glassy samples of elastic spheres, we first equilibrate the liquid phase at $T\!=\!0.004/k_B$ (here the computer $T_g\!\approx\!0.0017/k_B$), and follow the equilibration with an instantaneous quench using a nonlinear conjugate gradient algorithm \cite{macopt_cg} to form a glass. Following this protocol, we created and analyzed an ensemble of 1300 independent glassy samples, whose pressure-to-bulk modulus ratio is $p/K\!\approx\!0.173$.

\subsection{Stillinger-Weber network glass}

Originally developed to model silicon \cite{Stillinger_Weber}, the Stillinger-Weber (SW) potential has become widely used to model the phase behavior of tetrahedral liquids, from investigating liquid-liquid phase separation \cite{vasisht2011liquid}, to ice nucleation \cite{li2011homogeneous}. This model and its dynamics and thermodynamics are widespread in the literature; here we nevertheless spell out the potential energy $U_{\mbox{\tiny SW}}$ definition, followed by expressions for its Hessian matrix $\calBold{M}_{mn}\!\equiv\!\frac{\partial^2 U_{\mbox{\tiny SW}}}{\partial \xv_m \partial \xv_n}$, which, to the best of our knowledge, is not available in the current literature. 

The SW model is a monocomponent system of $N$ identical particles of mass $m$ whose interaction potential consists of both a short-ranged, two-body interaction 
\begin{equation}
\varphi_2(r_{ij}) =A\varepsilon\left(\frac{ B\sigma^4}{r_{ij}^4} - 1 \right) \exp{\left(\frac{\sigma}{r_{ij} - r_c}\right)}\,,
\end{equation}
and a three-body term 
\begin{multline}
\varphi_3(r_{ij},r_{ik},\theta_{jik}) = \lambda\varepsilon \left(\cos \theta_{jik} -\cos\theta_0 \right)^2 \times \\ \exp{\left(\frac{\gamma\sigma}{r_{ij} - r_c}\right)}\exp{\left(\frac{\gamma\sigma}{r_{ik} - r_c}\right)}\,,
\end{multline}
that favors triplets of atoms to form an angle $\theta_0\!\simeq\!109^{\mbox{\scriptsize o}}$. Here $\theta_{jik}$ is the angle formed by the $i,j$ and $i,k$ bonds, $r_{ij}\!\equiv\! |\xv_{ij}|$ is the pairwise distance between particles $i$ and $j$ (with $\xv_{ij}\!\equiv\!\xv_j\!-\! \xv_i)$, $r_c$ is a cutoff distance, $A,B,\lambda$ and $\gamma$ are dimensionless parameters, and $\sigma,\varepsilon$ are microscopic length and energy scales, respectively. The total potential energy of the system is computed as
\begin{equation}
U_{\mbox{\tiny SW}}=\sum_{i}\sum_{j>i}\varphi_2(r_{ij})+\sum_{i}\sum_{j\ne i}\sum_{k>j}\varphi_3(r_{ij},r_{ik},\theta_{jik}).
\end{equation}

We chose all microscopic parameters ($A,B,\gamma,r_c$) to be the same as in the SW parametrization of silicon \cite{Stillinger_Weber}, except for $\lambda$ --- whose high values favor local tetragonal order --- which was set to $18.75$. This value is found optimal to achieve good glass forming ability \cite{russo2018glass}. Lengths are expressed in terms of $\sigma$, and  energies in terms of $\varepsilon$. Simulations are performed in the $NVT$ ensemble using the highly parallel LAMMPS package~\cite{plimpton1993fast}. The temperature is controlled using the Nos\'{e}-Hoover thermostat, the number density is set to $N/V\!=\!0.52\sigma^{-3}$ and the temperature is fixed to $T\!=\!0.1\varepsilon/k_B$, which is far above our estimate for the glass transition temperature $T_g\!\approx\!0.014\varepsilon/k_B$.

To prepare glassy states, we collect a set of uncorrelated equilibrium configurations and perform an instantaneous quench of each configuration by minimizing the potential energy using a nonlinear conjugate gradient algorithm \cite{macopt_cg}. With this protocol, we gathered $1300$ independent glassy samples, each composed of $N=4096$ particles.

\subsubsection*{Expressions for the Hessian matrix}
\label{SW_appendix}

For ease of notation in spelling out the expression for the Hessian matrix $\calBold{M}_{mn}\!\equiv\!\frac{\partial^2 U_{\mbox{\tiny SW}}}{\partial \xv_m \partial \xv_n}$, we denote pairs of particles with Greek indices, e.g.~$r_{ij}\!\equiv\! r_\alpha$, then the potential energy can be spelled out as
\begin{equation}
U_{\mbox{\tiny SW}}=\sum_{\alpha}\varphi_2(r_\alpha)+\sum_{\alpha,\beta}\varphi_3(r_{\alpha},r_{\beta},\theta_{\alpha\beta}),
\end{equation}
where the sum over $\alpha,\beta$ runs over couples of pairs of particles that form a triple by sharing a common particle (e.g.~$i,j$ and $i,k$ where $j\! \ne\! k$), and $\theta_{\alpha\beta}$ is the angle formed between $\xv_\alpha$ and $\xv_\beta$. 

With these notations the Hessian matrix can be decomposed into a two-body part
\begin{equation}
\calBold{M}_{mn}^{2{\rm body}}=\sum_\alpha  \Gamma_{mn \alpha} \left( \frac{\mathcal I - \hat{\xv}_\alpha \hat{\xv}_\alpha}{r_\alpha}\frac{\partial \varphi_2}{\partial r_\alpha}+\hat{\xv}_\alpha \hat{\xv}_\alpha  \frac{\partial^2 \varphi_2}{\partial r_\alpha^2}\right)\,,
\end{equation}
and a three-body part
\begin{multline}
\calBold{M}_{mn}^{3{\rm body}}=\sum_{\alpha,\beta}\big(  \Gamma_{ mn\beta} \mathcal D_{\beta, \alpha}  + \Gamma_{mn\alpha} \mathcal D_{\alpha,\beta} +\\ \Gamma_{n\alpha}\Gamma_{m\beta} \mathcal X_{\alpha, \beta} +\Gamma_{n\beta}\Gamma_{m\alpha} \mathcal X_{ \beta, \alpha}\big)\,,
\end{multline}
where ${\cal I}$ is the identity tensor, $\Gamma_{m \alpha}\!\equiv\! (\delta_{jm}\!-\!\delta_{im})$, $\Gamma_{mn \alpha}\!\equiv\!(\delta_{jm}\!-\!\delta_{im})(\delta_{jn}\!-\!\delta_{in})$, and using the notation $c_{\alpha\beta}\!\equiv\!\cos \theta_{\alpha\beta}$,  $\mathcal D_{\beta, \beta}$ and $\mathcal X_{\alpha, \beta}$ read
\begin{multline}
\mathcal D_{\alpha, \beta}\equiv\frac{\mathcal I - \hat{\xv}_\alpha \hat{\xv}_\alpha}{r_\alpha}\frac{\partial \varphi_3}{\partial r_\alpha} + \hat{\xv}_\alpha \hat{\xv}_\alpha \frac{\partial^2 \varphi_3 }{\partial r_\alpha^2} + \frac{\partial^2 c_{\alpha\beta}}{\partial \xv_\alpha\partial \xv_\alpha} \frac{\partial \varphi_3}{\partial c_{\alpha\beta}} +\\ \frac{\partial c_{\alpha\beta}}{\partial \xv_\alpha} \frac{\partial c_{\alpha\beta}}{\partial \xv_\alpha} \frac{\partial^2 \varphi_3}{\partial c_{\alpha\beta}^2} + \left(\hat{\xv}_\alpha \frac{\partial c_{\alpha\beta}}{\partial \xv_\alpha} + \frac{\partial c_{\alpha\beta}}{\partial \xv_\alpha} \hat{\xv}_\alpha\right) \frac{\partial^2 \varphi_3}{\partial r_\alpha \partial c_{\alpha\beta}}\,,
\end{multline}
and 
\begin{multline}
\mathcal X_{\alpha, \beta}\equiv\hat{\xv}_\alpha \hat{ \xv}_\beta \frac{\partial^2 \varphi_3}{\partial r_\alpha \partial r_\beta} + \frac{\partial c_{\alpha\beta}}{\partial \xv_\alpha} \hat{\xv}_\beta \frac{\partial^2 \varphi_3}{\partial r_\beta \partial c_{\alpha\beta}} +\\ \frac{\partial^2 c_{\alpha\beta}}{\partial \xv_\alpha \partial \xv_\beta} \frac{\partial \varphi_3}{\partial c_{\alpha\beta}} + \frac{\partial c_{\alpha\beta}}{\partial \xv_\alpha} \frac{\partial c_{\alpha\beta}}{\partial \xv_\beta} \frac{\partial^2 \varphi_3}{\partial c_{\alpha\beta}^2} +\hat{\xv}_\alpha \frac{\partial c_{\alpha\beta}}{\partial \xv_\beta}  \frac{\partial^2 \varphi_3}{\partial r_\alpha \partial c_{\alpha\beta}},
\end{multline}
respectively. Furthermore 
\begin{equation}
\frac{\partial c_{\alpha\beta} }{\partial \xv_\alpha} = \frac{\hat{\xv}_\beta - \hat{\xv}_\alpha c_{\alpha\beta}}{r_\alpha}\,,
\end{equation}
\begin{equation}
\frac{\partial^2 c_{\alpha\beta}}{\partial \xv_\alpha\partial \xv_\alpha} = \frac{1}{r_\alpha^2}\left( \left(3\hat{\xv}_\alpha\hat{\xv}_\alpha - \mathcal I \right)c_{\alpha\beta} -\left(\hat{\xv}_\alpha \hat{\xv}_\beta + \hat{\xv}_\beta \hat{\xv}_\alpha \right)\right)\,,
\end{equation}
and
\begin{equation}
\frac{\partial^2 c_{\alpha\beta}}{\partial \xv_\alpha \partial\xv_\beta}= \frac1{r_\alpha r_\beta} \left(\mathcal I - \hat{\xv}_\alpha \hat{\xv}_\alpha -  \hat{\xv}_\beta  \hat{\xv}_\beta +  \hat{\xv}_\alpha  \hat{\xv}_\beta c_{\alpha\beta} \right)
\end{equation}
The implementation of the Hessian was validated using finite differences.

\subsection{Molecular glass}
The well-known model by Lewis-Wahnstr\"om \cite{LEWIS1993295} for the fragile glass former ortho-terphenyl (OTP) describes the OTP molecule as a rigid triangular molecule with site interactions at each vertex of the triangle, where each of these sites represents a whole phenyl ring that interacts with sites of different molecules via a Lennard-Jones potential. Inspired by this description, we consider a system of $N_m$ molecules comprised of $N\!=\!3N_m$ particles in three dimensions. We model the OTP molecule as a three-site isosceles triangle with two sides of length $\sigma$, and an angle between them of 75$^{\rm o}$. The intermolecular (site-site) interactions are given by the same smoothed Lennard Jones pairwise potential $\varphi_{\mbox{\tiny LJ}}$ as employed in the polymer system, see Eq.~(\ref{smoothLJ}). Lengths and energies are expressed in terms of $\sigma$ and $\varepsilon$, respectively.

\begin{figure}[ht!]
  \includegraphics[scale=1]{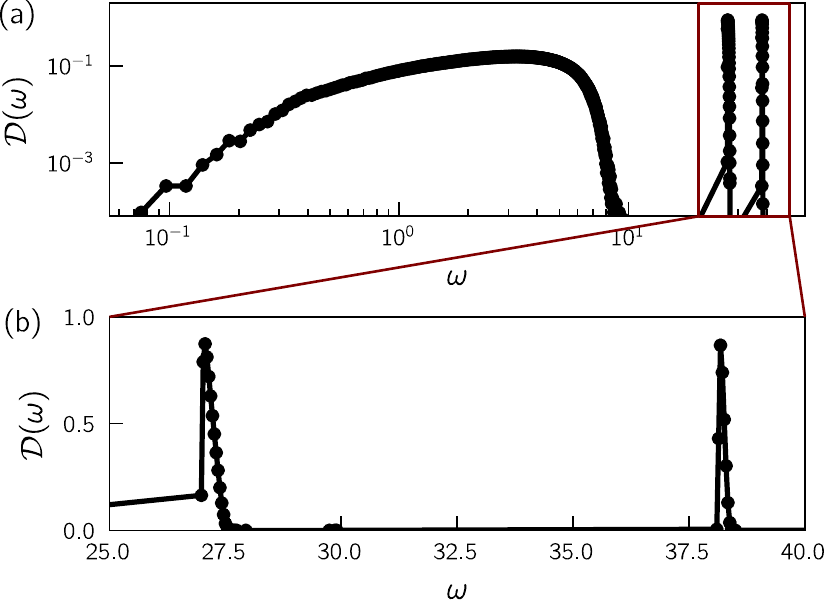}
  \caption{\footnotesize (a) Full density of states of the OTP model with $N=3000$. (b) Zoom on the two peaks associated to the intra-molecular vibrational modes.}
  \label{fig_ap:1}
\end{figure}

In contrast to \cite{LEWIS1993295}, in our model the three sites within a single molecule interact via a stiff harmonic potential, that reads:
\begin{equation}\label{harmonic}
\varphi_{\mbox{\tiny bonds}} =  \sFrac{1}{2} k (r_{ij} - l_0)^{2},
\end{equation}
where $r_{ij}$ is the distance between the $i^{\mbox{\tiny th}}$ and $j^{\mbox{\tiny th}}$ atoms in a molecule, $l_0$ represents the rest length of the intramolecular bonds, and $k\!=\!5\!\times\!10^3\varepsilon/\sigma^2$ denotes the stiffness of the intramolecular bonds, chosen to be roughly two orders of magnitude higher than typical intermolecular stiffnesses.

Glassy samples are prepared with the same method as employed in the polymeric system. We equilibrate our system in the {\it{NVT}} ensemble at a number density $N/V\!=\!0.80 \sigma^{-3}$ and a temperature $T\!=\!4.0\varepsilon/k_{B}$. Uncorrelated equilibrium configurations are prepared using conventional molecular dynamics and a Berendsen thermostat \cite{berendsen}, for which we set the time constant to $\tau_{_T}\!=\!1.0\sqrt{m\sigma^2/\varepsilon}$. After equilibration, an instantaneous quench to zero temperature using a conjugate gradient algorithm is performed. With this procedure we generate $5000$ independent glassy samples of $N_{m}\!=\!3000$ molecules that contains $N\!=\!9000$ atoms. In Fig.~\ref{fig_ap:1}, we provide the full density of states of the OTP system.

% Glassy samples are prepared with the same method as employed in the polymeric system. We equilibrate our system in the $NVT$ ensemble via Brownian dynamics at number density $N/V\!=\!0.8\sigma^{-3}$ and temperature $T\!=\!4.0\varepsilon/k_B$. \comment{With this procedure we generated $5000$ independent glassy samples of $N_m\!=\!3000$ molecules which contain $N\!=\!9000$ atoms. }

\subsection{Polymer glass}

Polymer melts are usually coarse-grained via simple bead-spring models where the actual monomer chemistry is replaced by an effective bead. Probably the most famous, the Kremer-Grest Model describes a polymeric chain via the finite extensible nonlinear elastic (FENE) potential\cite{kremer1990dynamics}. In this work, we adopt the same modeling expect that we replace the original repulsive Weeks-Chandler-Andersen potential \cite{weeks1971role} with a smoothed inverse-power law pairwise potential 
\begin{equation}\label{IPL}
\varphi_{\mbox{\tiny IPL}}(r_{ij}) = \left\{ \begin{array}{ccc}\!\!\varepsilon\left[ \left( \sFrac{\lambda}{r_{ij}} \right)^{10} \!\!+\! \sum\limits_{\ell=0}^3 c_{2\ell}\left(\sFrac{r_{ij}}{\lambda}\right)^{2\ell}\right]&,&r_{ij}\le r_c^{\mbox{\tiny IPL}}\\0&,&r_{ij}\ge r_c^{\mbox{\tiny IPL}}\end{array} \right.\!\!,
\end{equation}
where $\lambda$ is a microscopic length to be specified in what follows, and the coefficients $c_{2\ell}$ are determined by demanding that $\varphi_{\mbox{\tiny IPL}}$ vanishes continuously up to three derivatives at the cutoff $r_c^{\mbox{\tiny IPL}}$, see e.g.~\cite{cge_paper}.

The full modified FENE (mFENE) potential for nearest bonded monomers reads
\begin{equation}\label{FENE}
\varphi_{\mbox{\tiny mFENE}}(r_{ij}) = \varphi_{\mbox{\tiny IPL}}(r_{ij}) - \sFrac{1}{2}\kappa l_0^2 \ln\big(1-(r_{ij}/l_0)^2\big).
\end{equation}
Non-bonded intramolecular monomers only interact via $\varphi_{\mbox{\tiny IPL}}$ to account for volume exclusion. Intermolecular monomers interactions are given by a smoothed Lennard Jones pairwise potential (see e.g.~\cite{fsp}) of the form:
\begin{equation}\label{smoothLJ}
    \varphi_{\mbox{\tiny LJ}}(r_{ij}) \!=\!
\left\{
\begin{array}{cr}
\!\!6\varepsilon\!\! \left[ \big(\frac{\sigma}{r_{ij}}\big)^{12}\!\! - \big(\frac{\sigma}{r_{ij}}\big) ^{6}\! +\! \sum\limits_{l=0} ^{3} c_{2l} \big(\frac{r_{ij}}{\sigma}\big)^{2l} \right]\!,
     & \! r_{ij}\!<\! r_{c}^{\mbox{\tiny LJ}},  \\
0 , &\! r_{ij}\!\geq\! r_{c}^{\mbox{\tiny LJ}},
\end{array}
\right.
\end{equation}
where $r_{ij}$ is the distance between the $i^{\mbox{\tiny th}}$ and $j^{\mbox{\tiny th}}$ particles, $\varepsilon$ is microscopic energy scale, $\sigma$ is a microscopic length scale, and the coefficients $c_{2l}$ are determined by requiring that three derivatives of $\varphi_{\mbox{\tiny LJ}}$ with respect to $r_{ij}$ vanish continuously at the cutoff $r_{c}^{\mbox{\tiny LJ}}\!=\! 2\sigma$ in the same manner as done for $\varphi_{\mbox{\tiny IPL}}$. In practice, we set $\lambda\!=\!1.2\sigma$, $r_c^{\mbox{\tiny IPL}}\!=\!1.776\sigma$, $l_0=1.5\sigma$, and $\kappa=30.0\varepsilon/\sigma^2$.

To prepare the glassy samples, we first equilibrate the system in the {\it{NVT}} ensemble at a number density $N/V\!=\!0.80 \sigma^{-3}$ and a temperature $T\!=\!4.0\varepsilon/k_{B}$, the latter residing far above the glass transition temperature. To this aim we used molecular dynamics and we employed the Berendsen thermostat \cite{berendsen}, for which we set the time constant to $\tau_{_T}\!=\!1.0\sqrt{m\sigma^2/\varepsilon}$. After equilibration, the energy is minimized instantaneously using a standard conjugate gradient algorithm. Following this procedure, we generated $5000$ independent glassy samples of $N_{c} \!=\! 900$ chains composed of $10$ monomers for a total of $N\!=\!9000$ particles.

\subsection{CuZr bulk metallic glass (BMG)}
\label{BMG_appendix}

BMGs are simulated through the Embedded Atom Method (EAM)~\cite{cheng2009atomic,cheng2011atomic,BMG_site} in which the potential energy for atom $i$ is given by
\begin{equation}
E_i^{\mbox{\tiny BMG}} = F_\alpha \bigg(\sum_{j \neq i}\ \rho_\beta (r_{ij})\bigg) +
      \frac{1}{2} \sum_{j \neq i} \phi_{\alpha\beta} (r_{ij}) \ ,
\end{equation}
where the summations are over neighboring atoms $j$ within a cutoff, and $\alpha$ and $\beta$ are the element types of atoms $i$ and $j$ respectively. The values of the embedding function $F_\alpha$, the pair potential function $\phi$ and the effective charge density $\rho_\beta$ are derived from \textit{ab initio} calculations as well as from experimental data, and are provided in~\cite{BMG_site}.

We prepared glassy samples with $8000$ atoms out of which $46\%$ are Copper (Cu) and $54\%$ Zirconium (Zr). The number density is set to $N/V\!=\!0.058\AA^{-3}$ and the atom masses are set to their experimental values $91.22$ grams/mole for Zirconium and $63.54$ grams/mole for Copper.
We equilibrate the liquid phase at $1500$K and instantaneously quench by a conjugate-gradient algorithm. Our ensemble consists of 2086 samples with a shear modulus mean of $21.0$GPa. Simulations are performed in the $NVT$ ensemble using the highly parallel LAMMPS package~\cite{plimpton1993fast}.

The Hessian matrix for this model was calculated by moving each atom an infinitesimal distance $\Delta$ in each direction $x$, $y$ and $z$ and then measuring the change in forces $\calBold{F}$ on the atoms in the system. Following
\begin{equation}
\calBold{M}\cdot \xv = -\calBold{F} \ ,
\end{equation}
with e.g.~$\xv\!=\!\Delta \hat{\xv}$, for a displacement in the $x$ direction. We read out $\calBold{M}$ by normalizing the forces by the infinitesimal distance $\Delta$.

%\bibliography{references}

\begin{thebibliography}{55}%
\makeatletter
\providecommand \@ifxundefined [1]{%
 \@ifx{#1\undefined}
}%
\providecommand \@ifnum [1]{%
 \ifnum #1\expandafter \@firstoftwo
 \else \expandafter \@secondoftwo
 \fi
}%
\providecommand \@ifx [1]{%
 \ifx #1\expandafter \@firstoftwo
 \else \expandafter \@secondoftwo
 \fi
}%
\providecommand \natexlab [1]{#1}%
\providecommand \enquote  [1]{``#1''}%
\providecommand \bibnamefont  [1]{#1}%
\providecommand \bibfnamefont [1]{#1}%
\providecommand \citenamefont [1]{#1}%
\providecommand \href@noop [0]{\@secondoftwo}%
\providecommand \href [0]{\begingroup \@sanitize@url \@href}%
\providecommand \@href[1]{\@@startlink{#1}\@@href}%
\providecommand \@@href[1]{\endgroup#1\@@endlink}%
\providecommand \@sanitize@url [0]{\catcode `\\12\catcode `\$12\catcode
  `\&12\catcode `\#12\catcode `\^12\catcode `\_12\catcode `\%12\relax}%
\providecommand \@@startlink[1]{}%
\providecommand \@@endlink[0]{}%
\providecommand \url  [0]{\begingroup\@sanitize@url \@url }%
\providecommand \@url [1]{\endgroup\@href {#1}{\urlprefix }}%
\providecommand \urlprefix  [0]{URL }%
\providecommand \Eprint [0]{\href }%
\providecommand \doibase [0]{https://doi.org/}%
\providecommand \selectlanguage [0]{\@gobble}%
\providecommand \bibinfo  [0]{\@secondoftwo}%
\providecommand \bibfield  [0]{\@secondoftwo}%
\providecommand \translation [1]{[#1]}%
\providecommand \BibitemOpen [0]{}%
\providecommand \bibitemStop [0]{}%
\providecommand \bibitemNoStop [0]{.\EOS\space}%
\providecommand \EOS [0]{\spacefactor3000\relax}%
\providecommand \BibitemShut  [1]{\csname bibitem#1\endcsname}%
\let\auto@bib@innerbib\@empty
%</preamble>
\bibitem [{\citenamefont {Kittel}(2005)}]{kittel2005introduction}%
  \BibitemOpen
  \bibfield  {author} {\bibinfo {author} {\bibfnamefont {C.}~\bibnamefont
  {Kittel}},\ }\href@noop {} {\emph {\bibinfo {title} {Introduction to solid
  state physics}}}\ (\bibinfo  {publisher} {Wiley},\ \bibinfo {year}
  {2005})\BibitemShut {NoStop}%
\bibitem [{\citenamefont {Tsvelik}(2003)}]{tsvelik2003quantum}%
  \BibitemOpen
  \bibfield  {author} {\bibinfo {author} {\bibfnamefont {A.~M.}\ \bibnamefont
  {Tsvelik}},\ }\href@noop {} {\emph {\bibinfo {title} {Quantum field theory in
  condensed matter physics}}}\ (\bibinfo  {publisher} {Cambridge university
  press},\ \bibinfo {year} {2003})\BibitemShut {NoStop}%
\bibitem [{\citenamefont {Ketterson}(2016)}]{ketterson2016physics}%
  \BibitemOpen
  \bibfield  {author} {\bibinfo {author} {\bibfnamefont {J.~B.}\ \bibnamefont
  {Ketterson}},\ }\href {https://DOI:10.1093/acprof:oso/9780198742906.003.0017}
  {\emph {\bibinfo {title} {The Physics of solids}}}\ (\bibinfo  {publisher}
  {Oxford University Press},\ \bibinfo {year} {2016})\BibitemShut {NoStop}%
\bibitem [{\citenamefont {Suzuki}\ \emph {et~al.}(2013)\citenamefont {Suzuki},
  \citenamefont {Takeuchi},\ and\ \citenamefont
  {Yoshinaga}}]{suzuki2013dislocation}%
  \BibitemOpen
  \bibfield  {author} {\bibinfo {author} {\bibfnamefont {T.}~\bibnamefont
  {Suzuki}}, \bibinfo {author} {\bibfnamefont {S.}~\bibnamefont {Takeuchi}},\
  and\ \bibinfo {author} {\bibfnamefont {H.}~\bibnamefont {Yoshinaga}},\
  }\href@noop {} {\emph {\bibinfo {title} {Dislocation dynamics and
  plasticity}}},\ Vol.~\bibinfo {volume} {12}\ (\bibinfo  {publisher} {Springer
  Science \& Business Media},\ \bibinfo {year} {2013})\BibitemShut {NoStop}%
\bibitem [{\citenamefont {Zeller}\ and\ \citenamefont
  {Pohl}(1971)}]{Zeller_and_Pohl_prb_1971}%
  \BibitemOpen
  \bibfield  {author} {\bibinfo {author} {\bibfnamefont {R.~C.}\ \bibnamefont
  {Zeller}}\ and\ \bibinfo {author} {\bibfnamefont {R.~O.}\ \bibnamefont
  {Pohl}},\ }\bibfield  {title} {\bibinfo {title} {Thermal conductivity and
  specific heat of noncrystalline solids},\ }\href
  {https://doi.org/10.1103/PhysRevB.4.2029} {\bibfield  {journal} {\bibinfo
  {journal} {Phys. Rev. B}\ }\textbf {\bibinfo {volume} {4}},\ \bibinfo {pages}
  {2029} (\bibinfo {year} {1971})}\BibitemShut {NoStop}%
\bibitem [{\citenamefont {Anderson}\ \emph {et~al.}(1972)\citenamefont
  {Anderson}, \citenamefont {Halperin},\ and\ \citenamefont
  {Varma}}]{Anderson}%
  \BibitemOpen
  \bibfield  {author} {\bibinfo {author} {\bibfnamefont {P.~W.}\ \bibnamefont
  {Anderson}}, \bibinfo {author} {\bibfnamefont {B.~I.}\ \bibnamefont
  {Halperin}},\ and\ \bibinfo {author} {\bibfnamefont {C.~M.}\ \bibnamefont
  {Varma}},\ }\bibfield  {title} {\bibinfo {title} {Anomalous low-temperature
  thermal properties of glasses and spin glasses},\ }\href
  {https://doi.org/10.1080/14786437208229210} {\bibfield  {journal} {\bibinfo
  {journal} {Philos. Mag.}\ }\textbf {\bibinfo {volume} {25}},\ \bibinfo
  {pages} {1} (\bibinfo {year} {1972})}\BibitemShut {NoStop}%
\bibitem [{\citenamefont {Phillips}(1972)}]{Phillips}%
  \BibitemOpen
  \bibfield  {author} {\bibinfo {author} {\bibfnamefont {W.}~\bibnamefont
  {Phillips}},\ }\bibfield  {title} {\bibinfo {title} {Tunneling states in
  amorphous solids},\ }\href {https://doi.org/10.1007/BF00660072} {\bibfield
  {journal} {\bibinfo  {journal} {J. Low Temp. Phys.}\ }\textbf {\bibinfo
  {volume} {7}},\ \bibinfo {pages} {351} (\bibinfo {year} {1972})}\BibitemShut
  {NoStop}%
\bibitem [{\citenamefont {Falk}\ and\ \citenamefont
  {Langer}(1998)}]{falk_langer_stz}%
  \BibitemOpen
  \bibfield  {author} {\bibinfo {author} {\bibfnamefont {M.~L.}\ \bibnamefont
  {Falk}}\ and\ \bibinfo {author} {\bibfnamefont {J.~S.}\ \bibnamefont
  {Langer}},\ }\bibfield  {title} {\bibinfo {title} {Dynamics of viscoplastic
  deformation in amorphous solids},\ }\href
  {https://doi.org/10.1103/PhysRevE.57.7192} {\bibfield  {journal} {\bibinfo
  {journal} {Phys. Rev. E}\ }\textbf {\bibinfo {volume} {57}},\ \bibinfo
  {pages} {7192} (\bibinfo {year} {1998})}\BibitemShut {NoStop}%
\bibitem [{\citenamefont {Gartner}\ and\ \citenamefont
  {Lerner}(2016)}]{plastic_modes_prerc}%
  \BibitemOpen
  \bibfield  {author} {\bibinfo {author} {\bibfnamefont {L.}~\bibnamefont
  {Gartner}}\ and\ \bibinfo {author} {\bibfnamefont {E.}~\bibnamefont
  {Lerner}},\ }\bibfield  {title} {\bibinfo {title} {Nonlinear plastic modes in
  disordered solids},\ }\href {https://doi.org/10.1103/PhysRevE.93.011001}
  {\bibfield  {journal} {\bibinfo  {journal} {Phys. Rev. E}\ }\textbf {\bibinfo
  {volume} {93}},\ \bibinfo {pages} {011001} (\bibinfo {year}
  {2016})}\BibitemShut {NoStop}%
\bibitem [{\citenamefont {Buchenau}\ \emph {et~al.}(1991)\citenamefont
  {Buchenau}, \citenamefont {Galperin}, \citenamefont {Gurevich},\ and\
  \citenamefont {Schober}}]{soft_potential_model_1991}%
  \BibitemOpen
  \bibfield  {author} {\bibinfo {author} {\bibfnamefont {U.}~\bibnamefont
  {Buchenau}}, \bibinfo {author} {\bibfnamefont {Y.~M.}\ \bibnamefont
  {Galperin}}, \bibinfo {author} {\bibfnamefont {V.~L.}\ \bibnamefont
  {Gurevich}},\ and\ \bibinfo {author} {\bibfnamefont {H.~R.}\ \bibnamefont
  {Schober}},\ }\bibfield  {title} {\bibinfo {title} {Anharmonic potentials and
  vibrational localization in glasses},\ }\href
  {https://doi.org/10.1103/PhysRevB.43.5039} {\bibfield  {journal} {\bibinfo
  {journal} {Phys. Rev. B}\ }\textbf {\bibinfo {volume} {43}},\ \bibinfo
  {pages} {5039} (\bibinfo {year} {1991})}\BibitemShut {NoStop}%
\bibitem [{\citenamefont {Gurevich}\ \emph {et~al.}(2003)\citenamefont
  {Gurevich}, \citenamefont {Parshin},\ and\ \citenamefont
  {Schober}}]{Gurevich2003}%
  \BibitemOpen
  \bibfield  {author} {\bibinfo {author} {\bibfnamefont {V.~L.}\ \bibnamefont
  {Gurevich}}, \bibinfo {author} {\bibfnamefont {D.~A.}\ \bibnamefont
  {Parshin}},\ and\ \bibinfo {author} {\bibfnamefont {H.~R.}\ \bibnamefont
  {Schober}},\ }\bibfield  {title} {\bibinfo {title} {Anharmonicity,
  vibrational instability, and the boson peak in glasses},\ }\href
  {https://doi.org/10.1103/PhysRevB.67.094203} {\bibfield  {journal} {\bibinfo
  {journal} {Phys. Rev. B}\ }\textbf {\bibinfo {volume} {67}},\ \bibinfo
  {pages} {094203} (\bibinfo {year} {2003})}\BibitemShut {NoStop}%
\bibitem [{\citenamefont {Parshin}\ \emph {et~al.}(2007)\citenamefont
  {Parshin}, \citenamefont {Schober},\ and\ \citenamefont
  {Gurevich}}]{Gurevich2007}%
  \BibitemOpen
  \bibfield  {author} {\bibinfo {author} {\bibfnamefont {D.~A.}\ \bibnamefont
  {Parshin}}, \bibinfo {author} {\bibfnamefont {H.~R.}\ \bibnamefont
  {Schober}},\ and\ \bibinfo {author} {\bibfnamefont {V.~L.}\ \bibnamefont
  {Gurevich}},\ }\bibfield  {title} {\bibinfo {title} {Vibrational instability,
  two-level systems, and the boson peak in glasses},\ }\href
  {https://doi.org/10.1103/PhysRevB.76.064206} {\bibfield  {journal} {\bibinfo
  {journal} {Phys. Rev. B}\ }\textbf {\bibinfo {volume} {76}},\ \bibinfo
  {pages} {064206} (\bibinfo {year} {2007})}\BibitemShut {NoStop}%
\bibitem [{\citenamefont {Leonforte}\ \emph {et~al.}(2005)\citenamefont
  {Leonforte}, \citenamefont {Boissi\`ere}, \citenamefont {Tanguy},
  \citenamefont {Wittmer},\ and\ \citenamefont {Barrat}}]{barrat_3d}%
  \BibitemOpen
  \bibfield  {author} {\bibinfo {author} {\bibfnamefont {F.}~\bibnamefont
  {Leonforte}}, \bibinfo {author} {\bibfnamefont {R.}~\bibnamefont
  {Boissi\`ere}}, \bibinfo {author} {\bibfnamefont {A.}~\bibnamefont {Tanguy}},
  \bibinfo {author} {\bibfnamefont {J.~P.}\ \bibnamefont {Wittmer}},\ and\
  \bibinfo {author} {\bibfnamefont {J.-L.}\ \bibnamefont {Barrat}},\ }\bibfield
   {title} {\bibinfo {title} {Continuum limit of amorphous elastic bodies. iii.
  three-dimensional systems},\ }\href
  {https://doi.org/10.1103/PhysRevB.72.224206} {\bibfield  {journal} {\bibinfo
  {journal} {Phys. Rev. B}\ }\textbf {\bibinfo {volume} {72}},\ \bibinfo
  {pages} {224206} (\bibinfo {year} {2005})}\BibitemShut {NoStop}%
\bibitem [{\citenamefont {Schirmacher}\ \emph {et~al.}(2007)\citenamefont
  {Schirmacher}, \citenamefont {Ruocco},\ and\ \citenamefont
  {Scopigno}}]{Schirmacher_prl_2007}%
  \BibitemOpen
  \bibfield  {author} {\bibinfo {author} {\bibfnamefont {W.}~\bibnamefont
  {Schirmacher}}, \bibinfo {author} {\bibfnamefont {G.}~\bibnamefont
  {Ruocco}},\ and\ \bibinfo {author} {\bibfnamefont {T.}~\bibnamefont
  {Scopigno}},\ }\bibfield  {title} {\bibinfo {title} {Acoustic attenuation in
  glasses and its relation with the boson peak},\ }\href
  {https://doi.org/10.1103/PhysRevLett.98.025501} {\bibfield  {journal}
  {\bibinfo  {journal} {Phys. Rev. Lett.}\ }\textbf {\bibinfo {volume} {98}},\
  \bibinfo {pages} {025501} (\bibinfo {year} {2007})}\BibitemShut {NoStop}%
\bibitem [{\citenamefont {Baldi}\ \emph {et~al.}(2010)\citenamefont {Baldi},
  \citenamefont {Giordano}, \citenamefont {Monaco},\ and\ \citenamefont
  {Ruta}}]{experiments_1620K_vSiO2}%
  \BibitemOpen
  \bibfield  {author} {\bibinfo {author} {\bibfnamefont {G.}~\bibnamefont
  {Baldi}}, \bibinfo {author} {\bibfnamefont {V.~M.}\ \bibnamefont {Giordano}},
  \bibinfo {author} {\bibfnamefont {G.}~\bibnamefont {Monaco}},\ and\ \bibinfo
  {author} {\bibfnamefont {B.}~\bibnamefont {Ruta}},\ }\bibfield  {title}
  {\bibinfo {title} {Sound attenuation at terahertz frequencies and the boson
  peak of vitreous silica},\ }\href
  {https://doi.org/10.1103/PhysRevLett.104.195501} {\bibfield  {journal}
  {\bibinfo  {journal} {Phys. Rev. Lett.}\ }\textbf {\bibinfo {volume} {104}},\
  \bibinfo {pages} {195501} (\bibinfo {year} {2010})}\BibitemShut {NoStop}%
\bibitem [{\citenamefont {Wyart}(2010)}]{mw_EM_epl}%
  \BibitemOpen
  \bibfield  {author} {\bibinfo {author} {\bibfnamefont {M.}~\bibnamefont
  {Wyart}},\ }\bibfield  {title} {\bibinfo {title} {Scaling of phononic
  transport with connectivity in amorphous solids},\ }\href
  {http://stacks.iop.org/0295-5075/89/i=6/a=64001} {\bibfield  {journal}
  {\bibinfo  {journal} {Europhys. Lett.}\ }\textbf {\bibinfo {volume} {89}},\
  \bibinfo {pages} {64001} (\bibinfo {year} {2010})}\BibitemShut {NoStop}%
\bibitem [{\citenamefont {Hong}\ \emph {et~al.}(2011)\citenamefont {Hong},
  \citenamefont {Novikov},\ and\ \citenamefont
  {Sokolov}}]{sokolov_boson_peak_scale}%
  \BibitemOpen
  \bibfield  {author} {\bibinfo {author} {\bibfnamefont {L.}~\bibnamefont
  {Hong}}, \bibinfo {author} {\bibfnamefont {V.~N.}\ \bibnamefont {Novikov}},\
  and\ \bibinfo {author} {\bibfnamefont {A.~P.}\ \bibnamefont {Sokolov}},\
  }\bibfield  {title} {\bibinfo {title} {Dynamic heterogeneities, boson peak,
  and activation volume in glass-forming liquids},\ }\href
  {https://doi.org/10.1103/PhysRevE.83.061508} {\bibfield  {journal} {\bibinfo
  {journal} {Phys. Rev. E}\ }\textbf {\bibinfo {volume} {83}},\ \bibinfo
  {pages} {061508} (\bibinfo {year} {2011})}\BibitemShut {NoStop}%
\bibitem [{\citenamefont {Chumakov}\ \emph {et~al.}(2011)\citenamefont
  {Chumakov}, \citenamefont {Monaco}, \citenamefont {Monaco}, \citenamefont
  {Crichton}, \citenamefont {Bosak}, \citenamefont {R\"uffer}, \citenamefont
  {Meyer}, \citenamefont {Kargl}, \citenamefont {Comez}, \citenamefont
  {Fioretto}, \citenamefont {Giefers}, \citenamefont {Roitsch}, \citenamefont
  {Wortmann}, \citenamefont {Manghnani}, \citenamefont {Hushur}, \citenamefont
  {Williams}, \citenamefont {Balogh}, \citenamefont
  {Parli\ifmmode~\acute{n}\else \'{n}\fi{}ski}, \citenamefont {Jochym},\ and\
  \citenamefont {Piekarz}}]{Monaco_prl_2011}%
  \BibitemOpen
  \bibfield  {author} {\bibinfo {author} {\bibfnamefont {A.~I.}\ \bibnamefont
  {Chumakov}}, \bibinfo {author} {\bibfnamefont {G.}~\bibnamefont {Monaco}},
  \bibinfo {author} {\bibfnamefont {A.}~\bibnamefont {Monaco}}, \bibinfo
  {author} {\bibfnamefont {W.~A.}\ \bibnamefont {Crichton}}, \bibinfo {author}
  {\bibfnamefont {A.}~\bibnamefont {Bosak}}, \bibinfo {author} {\bibfnamefont
  {R.}~\bibnamefont {R\"uffer}}, \bibinfo {author} {\bibfnamefont
  {A.}~\bibnamefont {Meyer}}, \bibinfo {author} {\bibfnamefont
  {F.}~\bibnamefont {Kargl}}, \bibinfo {author} {\bibfnamefont
  {L.}~\bibnamefont {Comez}}, \bibinfo {author} {\bibfnamefont
  {D.}~\bibnamefont {Fioretto}}, \bibinfo {author} {\bibfnamefont
  {H.}~\bibnamefont {Giefers}}, \bibinfo {author} {\bibfnamefont
  {S.}~\bibnamefont {Roitsch}}, \bibinfo {author} {\bibfnamefont
  {G.}~\bibnamefont {Wortmann}}, \bibinfo {author} {\bibfnamefont {M.~H.}\
  \bibnamefont {Manghnani}}, \bibinfo {author} {\bibfnamefont {A.}~\bibnamefont
  {Hushur}}, \bibinfo {author} {\bibfnamefont {Q.}~\bibnamefont {Williams}},
  \bibinfo {author} {\bibfnamefont {J.}~\bibnamefont {Balogh}}, \bibinfo
  {author} {\bibfnamefont {K.}~\bibnamefont {Parli\ifmmode~\acute{n}\else
  \'{n}\fi{}ski}}, \bibinfo {author} {\bibfnamefont {P.}~\bibnamefont
  {Jochym}},\ and\ \bibinfo {author} {\bibfnamefont {P.}~\bibnamefont
  {Piekarz}},\ }\bibfield  {title} {\bibinfo {title} {Equivalence of the boson
  peak in glasses to the transverse acoustic van hove singularity in
  crystals},\ }\href {https://doi.org/10.1103/PhysRevLett.106.225501}
  {\bibfield  {journal} {\bibinfo  {journal} {Phys. Rev. Lett.}\ }\textbf
  {\bibinfo {volume} {106}},\ \bibinfo {pages} {225501} (\bibinfo {year}
  {2011})}\BibitemShut {NoStop}%
\bibitem [{\citenamefont {DeGiuli}\ \emph
  {et~al.}(2014{\natexlab{a}})\citenamefont {DeGiuli}, \citenamefont
  {Laversanne-Finot}, \citenamefont {During}, \citenamefont {Lerner},\ and\
  \citenamefont {Wyart}}]{eric_boson_peak_emt}%
  \BibitemOpen
  \bibfield  {author} {\bibinfo {author} {\bibfnamefont {E.}~\bibnamefont
  {DeGiuli}}, \bibinfo {author} {\bibfnamefont {A.}~\bibnamefont
  {Laversanne-Finot}}, \bibinfo {author} {\bibfnamefont {G.}~\bibnamefont
  {During}}, \bibinfo {author} {\bibfnamefont {E.}~\bibnamefont {Lerner}},\
  and\ \bibinfo {author} {\bibfnamefont {M.}~\bibnamefont {Wyart}},\ }\bibfield
   {title} {\bibinfo {title} {Effects of coordination and pressure on sound
  attenuation{,} boson peak and elasticity in amorphous solids},\ }\href
  {https://doi.org/10.1039/C4SM00561A} {\bibfield  {journal} {\bibinfo
  {journal} {Soft Matter}\ }\textbf {\bibinfo {volume} {10}},\ \bibinfo {pages}
  {5628} (\bibinfo {year} {2014}{\natexlab{a}})}\BibitemShut {NoStop}%
\bibitem [{\citenamefont {DeGiuli}\ \emph
  {et~al.}(2014{\natexlab{b}})\citenamefont {DeGiuli}, \citenamefont {Lerner},
  \citenamefont {Brito},\ and\ \citenamefont {Wyart}}]{eric_hard_spheres_emt}%
  \BibitemOpen
  \bibfield  {author} {\bibinfo {author} {\bibfnamefont {E.}~\bibnamefont
  {DeGiuli}}, \bibinfo {author} {\bibfnamefont {E.}~\bibnamefont {Lerner}},
  \bibinfo {author} {\bibfnamefont {C.}~\bibnamefont {Brito}},\ and\ \bibinfo
  {author} {\bibfnamefont {M.}~\bibnamefont {Wyart}},\ }\bibfield  {title}
  {\bibinfo {title} {Force distribution affects vibrational properties in
  hard-sphere glasses},\ }\href {https://doi.org/10.1073/pnas.1415298111}
  {\bibfield  {journal} {\bibinfo  {journal} {Proc. Natl. Acad. Sci. U.S.A.}\
  }\textbf {\bibinfo {volume} {111}},\ \bibinfo {pages} {17054} (\bibinfo
  {year} {2014}{\natexlab{b}})}\BibitemShut {NoStop}%
\bibitem [{\citenamefont {Franz}\ \emph {et~al.}(2015)\citenamefont {Franz},
  \citenamefont {Parisi}, \citenamefont {Urbani},\ and\ \citenamefont
  {Zamponi}}]{silvio}%
  \BibitemOpen
  \bibfield  {author} {\bibinfo {author} {\bibfnamefont {S.}~\bibnamefont
  {Franz}}, \bibinfo {author} {\bibfnamefont {G.}~\bibnamefont {Parisi}},
  \bibinfo {author} {\bibfnamefont {P.}~\bibnamefont {Urbani}},\ and\ \bibinfo
  {author} {\bibfnamefont {F.}~\bibnamefont {Zamponi}},\ }\bibfield  {title}
  {\bibinfo {title} {Universal spectrum of normal modes in low-temperature
  glasses},\ }\href {https://doi.org/10.1073/pnas.1511134112} {\bibfield
  {journal} {\bibinfo  {journal} {Proc. Natl. Acad. Sci. U.S.A.}\ }\textbf
  {\bibinfo {volume} {112}},\ \bibinfo {pages} {14539} (\bibinfo {year}
  {2015})}\BibitemShut {NoStop}%
\bibitem [{\citenamefont {Lerner}\ \emph {et~al.}(2016)\citenamefont {Lerner},
  \citenamefont {D\"uring},\ and\ \citenamefont
  {Bouchbinder}}]{modes_prl_2016}%
  \BibitemOpen
  \bibfield  {author} {\bibinfo {author} {\bibfnamefont {E.}~\bibnamefont
  {Lerner}}, \bibinfo {author} {\bibfnamefont {G.}~\bibnamefont {D\"uring}},\
  and\ \bibinfo {author} {\bibfnamefont {E.}~\bibnamefont {Bouchbinder}},\
  }\bibfield  {title} {\bibinfo {title} {Statistics and properties of
  low-frequency vibrational modes in structural glasses},\ }\href
  {https://doi.org/10.1103/PhysRevLett.117.035501} {\bibfield  {journal}
  {\bibinfo  {journal} {Phys. Rev. Lett.}\ }\textbf {\bibinfo {volume} {117}},\
  \bibinfo {pages} {035501} (\bibinfo {year} {2016})}\BibitemShut {NoStop}%
\bibitem [{\citenamefont {Kapteijns}\ \emph {et~al.}(2018)\citenamefont
  {Kapteijns}, \citenamefont {Bouchbinder},\ and\ \citenamefont
  {Lerner}}]{modes_prl_2018}%
  \BibitemOpen
  \bibfield  {author} {\bibinfo {author} {\bibfnamefont {G.}~\bibnamefont
  {Kapteijns}}, \bibinfo {author} {\bibfnamefont {E.}~\bibnamefont
  {Bouchbinder}},\ and\ \bibinfo {author} {\bibfnamefont {E.}~\bibnamefont
  {Lerner}},\ }\bibfield  {title} {\bibinfo {title} {Universal nonphononic
  density of states in 2d, 3d, and 4d glasses},\ }\href
  {https://doi.org/10.1103/PhysRevLett.121.055501} {\bibfield  {journal}
  {\bibinfo  {journal} {Phys. Rev. Lett.}\ }\textbf {\bibinfo {volume} {121}},\
  \bibinfo {pages} {055501} (\bibinfo {year} {2018})}\BibitemShut {NoStop}%
\bibitem [{\citenamefont {Lerner}\ and\ \citenamefont
  {Bouchbinder}(2018)}]{cge_paper}%
  \BibitemOpen
  \bibfield  {author} {\bibinfo {author} {\bibfnamefont {E.}~\bibnamefont
  {Lerner}}\ and\ \bibinfo {author} {\bibfnamefont {E.}~\bibnamefont
  {Bouchbinder}},\ }\bibfield  {title} {\bibinfo {title} {A characteristic
  energy scale in glasses},\ }\href {https://doi.org/10.1063/1.5024776}
  {\bibfield  {journal} {\bibinfo  {journal} {J. Chem. Phys.}\ }\textbf
  {\bibinfo {volume} {148}},\ \bibinfo {pages} {214502} (\bibinfo {year}
  {2018})}\BibitemShut {NoStop}%
\bibitem [{\citenamefont {Rainone}\ \emph {et~al.}(2020)\citenamefont
  {Rainone}, \citenamefont {Bouchbinder},\ and\ \citenamefont
  {Lerner}}]{pinching_pnas}%
  \BibitemOpen
  \bibfield  {author} {\bibinfo {author} {\bibfnamefont {C.}~\bibnamefont
  {Rainone}}, \bibinfo {author} {\bibfnamefont {E.}~\bibnamefont
  {Bouchbinder}},\ and\ \bibinfo {author} {\bibfnamefont {E.}~\bibnamefont
  {Lerner}},\ }\bibfield  {title} {\bibinfo {title} {Pinching a glass reveals
  key properties of its soft spots},\ }\href
  {https://doi.org/10.1073/pnas.1919958117} {\bibfield  {journal} {\bibinfo
  {journal} {Proc. Natl. Acad. Sci. U.S.A.}\ }\textbf {\bibinfo {volume}
  {117}},\ \bibinfo {pages} {5228} (\bibinfo {year} {2020})}\BibitemShut
  {NoStop}%
\bibitem [{\citenamefont {Mizuno}\ \emph {et~al.}(2017)\citenamefont {Mizuno},
  \citenamefont {Shiba},\ and\ \citenamefont {Ikeda}}]{ikeda_pnas}%
  \BibitemOpen
  \bibfield  {author} {\bibinfo {author} {\bibfnamefont {H.}~\bibnamefont
  {Mizuno}}, \bibinfo {author} {\bibfnamefont {H.}~\bibnamefont {Shiba}},\ and\
  \bibinfo {author} {\bibfnamefont {A.}~\bibnamefont {Ikeda}},\ }\bibfield
  {title} {\bibinfo {title} {Continuum limit of the vibrational properties of
  amorphous solids},\ }\href {https://doi.org/10.1073/pnas.1709015114}
  {\bibfield  {journal} {\bibinfo  {journal} {Proc. Natl. Acad. Sci. U.S.A.}\
  }\textbf {\bibinfo {volume} {114}},\ \bibinfo {pages} {E9767} (\bibinfo
  {year} {2017})}\BibitemShut {NoStop}%
\bibitem [{\citenamefont {Wang}\ \emph {et~al.}(2019)\citenamefont {Wang},
  \citenamefont {Ninarello}, \citenamefont {Guan}, \citenamefont {Berthier},
  \citenamefont {Szamel},\ and\ \citenamefont {Flenner}}]{LB_modes_2019}%
  \BibitemOpen
  \bibfield  {author} {\bibinfo {author} {\bibfnamefont {L.}~\bibnamefont
  {Wang}}, \bibinfo {author} {\bibfnamefont {A.}~\bibnamefont {Ninarello}},
  \bibinfo {author} {\bibfnamefont {P.}~\bibnamefont {Guan}}, \bibinfo {author}
  {\bibfnamefont {L.}~\bibnamefont {Berthier}}, \bibinfo {author}
  {\bibfnamefont {G.}~\bibnamefont {Szamel}},\ and\ \bibinfo {author}
  {\bibfnamefont {E.}~\bibnamefont {Flenner}},\ }\bibfield  {title} {\bibinfo
  {title} {Low-frequency vibrational modes of stable glasses},\ }\href
  {https://doi.org/10.1038/s41467-018-07978-1} {\bibfield  {journal} {\bibinfo
  {journal} {Nat. Commun.}\ }\textbf {\bibinfo {volume} {10}},\ \bibinfo
  {pages} {26} (\bibinfo {year} {2019})}\BibitemShut {NoStop}%
\bibitem [{\citenamefont {Shimada}\ \emph {et~al.}(2018)\citenamefont
  {Shimada}, \citenamefont {Mizuno}, \citenamefont {Wyart},\ and\ \citenamefont
  {Ikeda}}]{atsushi_core_size_pre}%
  \BibitemOpen
  \bibfield  {author} {\bibinfo {author} {\bibfnamefont {M.}~\bibnamefont
  {Shimada}}, \bibinfo {author} {\bibfnamefont {H.}~\bibnamefont {Mizuno}},
  \bibinfo {author} {\bibfnamefont {M.}~\bibnamefont {Wyart}},\ and\ \bibinfo
  {author} {\bibfnamefont {A.}~\bibnamefont {Ikeda}},\ }\bibfield  {title}
  {\bibinfo {title} {Spatial structure of quasilocalized vibrations in nearly
  jammed amorphous solids},\ }\href
  {https://doi.org/10.1103/PhysRevE.98.060901} {\bibfield  {journal} {\bibinfo
  {journal} {Phys. Rev. E}\ }\textbf {\bibinfo {volume} {98}},\ \bibinfo
  {pages} {060901} (\bibinfo {year} {2018})}\BibitemShut {NoStop}%
\bibitem [{\citenamefont {Bonfanti}\ \emph {et~al.}(2020)\citenamefont
  {Bonfanti}, \citenamefont {Guerra}, \citenamefont {Mondal}, \citenamefont
  {Procaccia},\ and\ \citenamefont {Zapperi}}]{bonfanti2020universal}%
  \BibitemOpen
  \bibfield  {author} {\bibinfo {author} {\bibfnamefont {S.}~\bibnamefont
  {Bonfanti}}, \bibinfo {author} {\bibfnamefont {R.}~\bibnamefont {Guerra}},
  \bibinfo {author} {\bibfnamefont {C.}~\bibnamefont {Mondal}}, \bibinfo
  {author} {\bibfnamefont {I.}~\bibnamefont {Procaccia}},\ and\ \bibinfo
  {author} {\bibfnamefont {S.}~\bibnamefont {Zapperi}},\ }\bibfield  {title}
  {\bibinfo {title} {Universal low-frequency vibrational modes in silica
  glasses},\ }\href {https://arxiv.org/abs/2003.07614} {\bibfield  {journal}
  {\bibinfo  {journal} {arXiv preprint arXiv:2003.07614}\ } (\bibinfo {year}
  {2020})}\BibitemShut {NoStop}%
\bibitem [{\citenamefont {Hertz}(2006)}]{hertz2006beruhrung}%
  \BibitemOpen
  \bibfield  {author} {\bibinfo {author} {\bibfnamefont {H.~R.}\ \bibnamefont
  {Hertz}},\ }\href@noop {} {\emph {\bibinfo {title} {{\"U}ber die
  Ber{\"u}hrung fester elastischer K{\"o}rper und {\"u}ber die H{\"a}rte}}}\
  (\bibinfo  {publisher} {Universit{\"a}tsbibliothek Johann Christian
  Senckenberg},\ \bibinfo {year} {2006})\BibitemShut {NoStop}%
\bibitem [{\citenamefont {O'Hern}\ \emph {et~al.}(2003)\citenamefont {O'Hern},
  \citenamefont {Silbert}, \citenamefont {Liu},\ and\ \citenamefont
  {Nagel}}]{ohern2003}%
  \BibitemOpen
  \bibfield  {author} {\bibinfo {author} {\bibfnamefont {C.~S.}\ \bibnamefont
  {O'Hern}}, \bibinfo {author} {\bibfnamefont {L.~E.}\ \bibnamefont {Silbert}},
  \bibinfo {author} {\bibfnamefont {A.~J.}\ \bibnamefont {Liu}},\ and\ \bibinfo
  {author} {\bibfnamefont {S.~R.}\ \bibnamefont {Nagel}},\ }\bibfield  {title}
  {\bibinfo {title} {Jamming at zero temperature and zero applied stress: The
  epitome of disorder},\ }\href {https://doi.org/10.1103/PhysRevE.68.011306}
  {\bibfield  {journal} {\bibinfo  {journal} {Phys. Rev. E}\ }\textbf {\bibinfo
  {volume} {68}},\ \bibinfo {pages} {011306} (\bibinfo {year}
  {2003})}\BibitemShut {NoStop}%
\bibitem [{\citenamefont {Liu}\ and\ \citenamefont {Nagel}(2010)}]{liu_review}%
  \BibitemOpen
  \bibfield  {author} {\bibinfo {author} {\bibfnamefont {A.~J.}\ \bibnamefont
  {Liu}}\ and\ \bibinfo {author} {\bibfnamefont {S.~R.}\ \bibnamefont
  {Nagel}},\ }\bibfield  {title} {\bibinfo {title} {The jamming transition and
  the marginally jammed solid},\ }\href
  {https://doi.org/10.1146/annurev-conmatphys-070909-104045} {\bibfield
  {journal} {\bibinfo  {journal} {Annu. Rev. Condens. Matter Phys.}\ }\textbf
  {\bibinfo {volume} {1}},\ \bibinfo {pages} {347} (\bibinfo {year}
  {2010})}\BibitemShut {NoStop}%
\bibitem [{\citenamefont {van Hecke}(2010)}]{van_hecke_review}%
  \BibitemOpen
  \bibfield  {author} {\bibinfo {author} {\bibfnamefont {M.}~\bibnamefont {van
  Hecke}},\ }\bibfield  {title} {\bibinfo {title} {Jamming of soft particles:
  geometry, mechanics, scaling and isostaticity},\ }\href
  {http://stacks.iop.org/0953-8984/22/i=3/a=033101} {\bibfield  {journal}
  {\bibinfo  {journal} {J. Phys.: Condens. Matter}\ }\textbf {\bibinfo {volume}
  {22}},\ \bibinfo {pages} {033101} (\bibinfo {year} {2010})}\BibitemShut
  {NoStop}%
\bibitem [{\citenamefont {Stillinger}\ and\ \citenamefont
  {Weber}(1985)}]{Stillinger_Weber}%
  \BibitemOpen
  \bibfield  {author} {\bibinfo {author} {\bibfnamefont {F.~H.}\ \bibnamefont
  {Stillinger}}\ and\ \bibinfo {author} {\bibfnamefont {T.~A.}\ \bibnamefont
  {Weber}},\ }\bibfield  {title} {\bibinfo {title} {Computer simulation of
  local order in condensed phases of silicon},\ }\href
  {https://doi.org/10.1103/PhysRevB.31.5262} {\bibfield  {journal} {\bibinfo
  {journal} {Phys. Rev. B}\ }\textbf {\bibinfo {volume} {31}},\ \bibinfo
  {pages} {5262} (\bibinfo {year} {1985})}\BibitemShut {NoStop}%
\bibitem [{\citenamefont {Molinero}\ and\ \citenamefont
  {Moore}(2009)}]{sw_vary_lambda}%
  \BibitemOpen
  \bibfield  {author} {\bibinfo {author} {\bibfnamefont {V.}~\bibnamefont
  {Molinero}}\ and\ \bibinfo {author} {\bibfnamefont {E.~B.}\ \bibnamefont
  {Moore}},\ }\bibfield  {title} {\bibinfo {title} {Water modeled as an
  intermediate element between carbon and silicon},\ }\href
  {https://doi.org/10.1021/jp805227c} {\bibfield  {journal} {\bibinfo
  {journal} {J. Phys. Chem. B}\ }\textbf {\bibinfo {volume} {113}},\ \bibinfo
  {pages} {4008} (\bibinfo {year} {2009})},\ \bibinfo {note} {pMID:
  18956896}\BibitemShut {NoStop}%
\bibitem [{\citenamefont {Mossa}\ \emph {et~al.}(2002)\citenamefont {Mossa},
  \citenamefont {La~Nave}, \citenamefont {Stanley}, \citenamefont {Donati},
  \citenamefont {Sciortino},\ and\ \citenamefont {Tartaglia}}]{otp_paper}%
  \BibitemOpen
  \bibfield  {author} {\bibinfo {author} {\bibfnamefont {S.}~\bibnamefont
  {Mossa}}, \bibinfo {author} {\bibfnamefont {E.}~\bibnamefont {La~Nave}},
  \bibinfo {author} {\bibfnamefont {H.~E.}\ \bibnamefont {Stanley}}, \bibinfo
  {author} {\bibfnamefont {C.}~\bibnamefont {Donati}}, \bibinfo {author}
  {\bibfnamefont {F.}~\bibnamefont {Sciortino}},\ and\ \bibinfo {author}
  {\bibfnamefont {P.}~\bibnamefont {Tartaglia}},\ }\bibfield  {title} {\bibinfo
  {title} {Dynamics and configurational entropy in the lewis-wahnstr\"om model
  for supercooled orthoterphenyl},\ }\href
  {https://doi.org/10.1103/PhysRevE.65.041205} {\bibfield  {journal} {\bibinfo
  {journal} {Phys. Rev. E}\ }\textbf {\bibinfo {volume} {65}},\ \bibinfo
  {pages} {041205} (\bibinfo {year} {2002})}\BibitemShut {NoStop}%
\bibitem [{\citenamefont {Lewis}\ and\ \citenamefont
  {Wahnstr{\"{o}}m}(1993)}]{LEWIS1993295}%
  \BibitemOpen
  \bibfield  {author} {\bibinfo {author} {\bibfnamefont {L.~J.}\ \bibnamefont
  {Lewis}}\ and\ \bibinfo {author} {\bibfnamefont {G.}~\bibnamefont
  {Wahnstr{\"{o}}m}},\ }\bibfield  {title} {\bibinfo {title} {Relaxation of a
  molecular glass at intermediate times},\ }\href
  {https://doi.org/https://doi.org/10.1016/0038-1098(93)90376-X} {\bibfield
  {journal} {\bibinfo  {journal} {Solid State Commun.}\ }\textbf {\bibinfo
  {volume} {86}},\ \bibinfo {pages} {295} (\bibinfo {year} {1993})}\BibitemShut
  {NoStop}%
\bibitem [{\citenamefont {Starr}\ \emph {et~al.}(2002)\citenamefont {Starr},
  \citenamefont {Schr{\o}der},\ and\ \citenamefont {Glotzer}}]{Starr2002}%
  \BibitemOpen
  \bibfield  {author} {\bibinfo {author} {\bibfnamefont {F.~W.}\ \bibnamefont
  {Starr}}, \bibinfo {author} {\bibfnamefont {T.~B.}\ \bibnamefont
  {Schr{\o}der}},\ and\ \bibinfo {author} {\bibfnamefont {S.~C.}\ \bibnamefont
  {Glotzer}},\ }\bibfield  {title} {\bibinfo {title} {Molecular dynamics
  simulation of a polymer melt with a nanoscopic particle},\ }\href
  {https://doi.org/10.1021/ma010626p} {\bibfield  {journal} {\bibinfo
  {journal} {Macromolecules}\ }\textbf {\bibinfo {volume} {35}},\ \bibinfo
  {pages} {4481} (\bibinfo {year} {2002})}\BibitemShut {NoStop}%
\bibitem [{\citenamefont {Kapteijns}\ \emph {et~al.}(2019)\citenamefont
  {Kapteijns}, \citenamefont {Ji}, \citenamefont {Brito}, \citenamefont
  {Wyart},\ and\ \citenamefont {Lerner}}]{fsp}%
  \BibitemOpen
  \bibfield  {author} {\bibinfo {author} {\bibfnamefont {G.}~\bibnamefont
  {Kapteijns}}, \bibinfo {author} {\bibfnamefont {W.}~\bibnamefont {Ji}},
  \bibinfo {author} {\bibfnamefont {C.}~\bibnamefont {Brito}}, \bibinfo
  {author} {\bibfnamefont {M.}~\bibnamefont {Wyart}},\ and\ \bibinfo {author}
  {\bibfnamefont {E.}~\bibnamefont {Lerner}},\ }\bibfield  {title} {\bibinfo
  {title} {Fast generation of ultrastable computer glasses by minimization of
  an augmented potential energy},\ }\href
  {https://doi.org/10.1103/PhysRevE.99.012106} {\bibfield  {journal} {\bibinfo
  {journal} {Phys. Rev. E}\ }\textbf {\bibinfo {volume} {99}},\ \bibinfo
  {pages} {012106} (\bibinfo {year} {2019})}\BibitemShut {NoStop}%
\bibitem [{\citenamefont {Cheng}\ \emph {et~al.}(2009)\citenamefont {Cheng},
  \citenamefont {Ma},\ and\ \citenamefont {Sheng}}]{cheng2009atomic}%
  \BibitemOpen
  \bibfield  {author} {\bibinfo {author} {\bibfnamefont {Y.~Q.}\ \bibnamefont
  {Cheng}}, \bibinfo {author} {\bibfnamefont {E.}~\bibnamefont {Ma}},\ and\
  \bibinfo {author} {\bibfnamefont {H.~W.}\ \bibnamefont {Sheng}},\ }\bibfield
  {title} {\bibinfo {title} {Atomic level structure in multicomponent bulk
  metallic glass},\ }\href {https://doi.org/10.1103/PhysRevLett.102.245501}
  {\bibfield  {journal} {\bibinfo  {journal} {Phys. Rev. Lett.}\ }\textbf
  {\bibinfo {volume} {102}},\ \bibinfo {pages} {245501} (\bibinfo {year}
  {2009})}\BibitemShut {NoStop}%
\bibitem [{\citenamefont {Cheng}\ and\ \citenamefont
  {Ma}(2011)}]{cheng2011atomic}%
  \BibitemOpen
  \bibfield  {author} {\bibinfo {author} {\bibfnamefont {Y.}~\bibnamefont
  {Cheng}}\ and\ \bibinfo {author} {\bibfnamefont {E.}~\bibnamefont {Ma}},\
  }\bibfield  {title} {\bibinfo {title} {Atomic-level structure and
  structure--property relationship in metallic glasses},\ }\href
  {https://www.sciencedirect.com/science/article/pii/S0079642510000691}
  {\bibfield  {journal} {\bibinfo  {journal} {Prog. Mater. Sci.}\ }\textbf
  {\bibinfo {volume} {56}},\ \bibinfo {pages} {379} (\bibinfo {year}
  {2011})}\BibitemShut {NoStop}%
\bibitem [{\citenamefont {Lerner}(2020)}]{finite_size_modes}%
  \BibitemOpen
  \bibfield  {author} {\bibinfo {author} {\bibfnamefont {E.}~\bibnamefont
  {Lerner}},\ }\bibfield  {title} {\bibinfo {title} {Finite-size effects in the
  nonphononic density of states in computer glasses},\ }\href
  {https://doi.org/10.1103/PhysRevE.101.032120} {\bibfield  {journal} {\bibinfo
   {journal} {Phys. Rev. E}\ }\textbf {\bibinfo {volume} {101}},\ \bibinfo
  {pages} {032120} (\bibinfo {year} {2020})}\BibitemShut {NoStop}%
\bibitem [{foo()}]{footnote}%
  \BibitemOpen
  \href@noop {} {\bibfield  {journal} {\bibinfo  {journal} {$e\!\sim\!1/N$ also
  holds for quasilocalized modes in three or more dimensions, as shown in
  e.g.~\cite{modes_prl_2016}}\ }}\BibitemShut {NoStop}%
\bibitem [{\citenamefont {Lerner}\ and\ \citenamefont
  {Bouchbinder}(2017)}]{protocol_prerc}%
  \BibitemOpen
  \bibfield  {author} {\bibinfo {author} {\bibfnamefont {E.}~\bibnamefont
  {Lerner}}\ and\ \bibinfo {author} {\bibfnamefont {E.}~\bibnamefont
  {Bouchbinder}},\ }\bibfield  {title} {\bibinfo {title} {Effect of
  instantaneous and continuous quenches on the density of vibrational modes in
  model glasses},\ }\href {https://doi.org/10.1103/PhysRevE.96.020104}
  {\bibfield  {journal} {\bibinfo  {journal} {Phys. Rev. E}\ }\textbf {\bibinfo
  {volume} {96}},\ \bibinfo {pages} {020104} (\bibinfo {year}
  {2017})}\BibitemShut {NoStop}%
\bibitem [{\citenamefont {Lutsko}(1989)}]{lutsko}%
  \BibitemOpen
  \bibfield  {author} {\bibinfo {author} {\bibfnamefont {J.~F.}\ \bibnamefont
  {Lutsko}},\ }\bibfield  {title} {\bibinfo {title} {Generalized expressions
  for the calculation of elastic constants by computer simulation},\ }\href
  {https://doi.org/10.1063/1.342716} {\bibfield  {journal} {\bibinfo  {journal}
  {J. Appl. Phys.}\ }\textbf {\bibinfo {volume} {65}},\ \bibinfo {pages} {2991}
  (\bibinfo {year} {1989})}\BibitemShut {NoStop}%
\bibitem [{\citenamefont {Lehoucq}\ \emph {et~al.}(1998)\citenamefont
  {Lehoucq}, \citenamefont {Sorensen},\ and\ \citenamefont {Yang}}]{arpack}%
  \BibitemOpen
  \bibfield  {author} {\bibinfo {author} {\bibfnamefont {R.~B.}\ \bibnamefont
  {Lehoucq}}, \bibinfo {author} {\bibfnamefont {D.~C.}\ \bibnamefont
  {Sorensen}},\ and\ \bibinfo {author} {\bibfnamefont {C.}~\bibnamefont
  {Yang}},\ }\href {https://doi.org/10.1137/1.9780898719628} {\emph {\bibinfo
  {title} {ARPACK Users' Guide}}}\ (\bibinfo  {publisher} {Society for
  Industrial and Applied Mathematics, Philadelphia},\ \bibinfo {year}
  {1998})\BibitemShut {NoStop}%
\bibitem [{\citenamefont {MacKay}(2004)}]{macopt_cg}%
  \BibitemOpen
  \bibfield  {author} {\bibinfo {author} {\bibfnamefont {D.}~\bibnamefont
  {MacKay}},\ }\href {http://www.inference.org.uk/mackay/c/macopt.html}
  {\bibinfo {title} {macopt optimizer}} (\bibinfo {year} {2004})\BibitemShut
  {NoStop}%
\bibitem [{\citenamefont {Vasisht}\ \emph {et~al.}(2011)\citenamefont
  {Vasisht}, \citenamefont {Saw},\ and\ \citenamefont
  {Sastry}}]{vasisht2011liquid}%
  \BibitemOpen
  \bibfield  {author} {\bibinfo {author} {\bibfnamefont {V.~V.}\ \bibnamefont
  {Vasisht}}, \bibinfo {author} {\bibfnamefont {S.}~\bibnamefont {Saw}},\ and\
  \bibinfo {author} {\bibfnamefont {S.}~\bibnamefont {Sastry}},\ }\bibfield
  {title} {\bibinfo {title} {Liquid--liquid critical point in supercooled
  silicon},\ }\href {https://doi.org/10.1038/nphys1993} {\bibfield  {journal}
  {\bibinfo  {journal} {Nat. Phys.}\ }\textbf {\bibinfo {volume} {7}},\
  \bibinfo {pages} {549} (\bibinfo {year} {2011})}\BibitemShut {NoStop}%
\bibitem [{\citenamefont {Li}\ \emph {et~al.}(2011)\citenamefont {Li},
  \citenamefont {Donadio}, \citenamefont {Russo},\ and\ \citenamefont
  {Galli}}]{li2011homogeneous}%
  \BibitemOpen
  \bibfield  {author} {\bibinfo {author} {\bibfnamefont {T.}~\bibnamefont
  {Li}}, \bibinfo {author} {\bibfnamefont {D.}~\bibnamefont {Donadio}},
  \bibinfo {author} {\bibfnamefont {G.}~\bibnamefont {Russo}},\ and\ \bibinfo
  {author} {\bibfnamefont {G.}~\bibnamefont {Galli}},\ }\bibfield  {title}
  {\bibinfo {title} {Homogeneous ice nucleation from supercooled water},\
  }\href {https://doi.org/10.1039/C1CP22167A} {\bibfield  {journal} {\bibinfo
  {journal} {Phys. Chem. Chem. Phys.}\ }\textbf {\bibinfo {volume} {13}},\
  \bibinfo {pages} {19807} (\bibinfo {year} {2011})}\BibitemShut {NoStop}%
\bibitem [{\citenamefont {Russo}\ \emph {et~al.}(2018)\citenamefont {Russo},
  \citenamefont {Romano},\ and\ \citenamefont {Tanaka}}]{russo2018glass}%
  \BibitemOpen
  \bibfield  {author} {\bibinfo {author} {\bibfnamefont {J.}~\bibnamefont
  {Russo}}, \bibinfo {author} {\bibfnamefont {F.}~\bibnamefont {Romano}},\ and\
  \bibinfo {author} {\bibfnamefont {H.}~\bibnamefont {Tanaka}},\ }\bibfield
  {title} {\bibinfo {title} {Glass forming ability in systems with competing
  orderings},\ }\href {https://doi.org/10.1103/PhysRevX.8.021040} {\bibfield
  {journal} {\bibinfo  {journal} {Phys. Rev. X}\ }\textbf {\bibinfo {volume}
  {8}},\ \bibinfo {pages} {021040} (\bibinfo {year} {2018})}\BibitemShut
  {NoStop}%
\bibitem [{\citenamefont {Plimpton}(1993)}]{plimpton1993fast}%
  \BibitemOpen
  \bibfield  {author} {\bibinfo {author} {\bibfnamefont {S.}~\bibnamefont
  {Plimpton}},\ }\href@noop {} {\emph {\bibinfo {title} {Fast parallel
  algorithms for short-range molecular dynamics}}},\ \bibinfo {type} {Tech.
  Rep.}\ (\bibinfo  {institution} {Sandia National Labs., Albuquerque, NM
  (United States)},\ \bibinfo {year} {1993})\BibitemShut {NoStop}%
\bibitem [{\citenamefont {Berendsen}\ \emph {et~al.}(1984)\citenamefont
  {Berendsen}, \citenamefont {Postma}, \citenamefont {van Gunsteren},
  \citenamefont {DiNola},\ and\ \citenamefont {Haak}}]{berendsen}%
  \BibitemOpen
  \bibfield  {author} {\bibinfo {author} {\bibfnamefont {H.~J.~C.}\
  \bibnamefont {Berendsen}}, \bibinfo {author} {\bibfnamefont {J.~P.~M.}\
  \bibnamefont {Postma}}, \bibinfo {author} {\bibfnamefont {W.~F.}\
  \bibnamefont {van Gunsteren}}, \bibinfo {author} {\bibfnamefont
  {A.}~\bibnamefont {DiNola}},\ and\ \bibinfo {author} {\bibfnamefont {J.~R.}\
  \bibnamefont {Haak}},\ }\bibfield  {title} {\bibinfo {title} {Molecular
  dynamics with coupling to an external bath},\ }\href
  {https://doi.org/http://dx.doi.org/10.1063/1.448118} {\bibfield  {journal}
  {\bibinfo  {journal} {J. Chem. Phys.}\ }\textbf {\bibinfo {volume} {81}},\
  \bibinfo {pages} {3684} (\bibinfo {year} {1984})}\BibitemShut {NoStop}%
\bibitem [{\citenamefont {Kremer}\ and\ \citenamefont
  {Grest}(1990)}]{kremer1990dynamics}%
  \BibitemOpen
  \bibfield  {author} {\bibinfo {author} {\bibfnamefont {K.}~\bibnamefont
  {Kremer}}\ and\ \bibinfo {author} {\bibfnamefont {G.~S.}\ \bibnamefont
  {Grest}},\ }\bibfield  {title} {\bibinfo {title} {Dynamics of entangled
  linear polymer melts:  a molecular‐dynamics simulation},\ }\href
  {https://doi.org/10.1063/1.458541} {\bibfield  {journal} {\bibinfo  {journal}
  {J. Chem. Phys.}\ }\textbf {\bibinfo {volume} {92}},\ \bibinfo {pages} {5057}
  (\bibinfo {year} {1990})}\BibitemShut {NoStop}%
\bibitem [{\citenamefont {Weeks}\ \emph {et~al.}(1971)\citenamefont {Weeks},
  \citenamefont {Chandler},\ and\ \citenamefont {Andersen}}]{weeks1971role}%
  \BibitemOpen
  \bibfield  {author} {\bibinfo {author} {\bibfnamefont {J.~D.}\ \bibnamefont
  {Weeks}}, \bibinfo {author} {\bibfnamefont {D.}~\bibnamefont {Chandler}},\
  and\ \bibinfo {author} {\bibfnamefont {H.~C.}\ \bibnamefont {Andersen}},\
  }\bibfield  {title} {\bibinfo {title} {Role of repulsive forces in
  determining the equilibrium structure of simple liquids},\ }\href
  {https://doi.org/10.1063/1.1674820} {\bibfield  {journal} {\bibinfo
  {journal} {J. Chem. Phys.}\ }\textbf {\bibinfo {volume} {54}},\ \bibinfo
  {pages} {5237} (\bibinfo {year} {1971})}\BibitemShut {NoStop}%
\bibitem [{BMG(2011)}]{BMG_site}%
  \BibitemOpen
  \href {https://sites.google.com/site/eampotentials/Home/CuZr} {\bibinfo
  {title} {Cu-zr eam potential}} (\bibinfo {year} {2011})\BibitemShut {NoStop}%
\end{thebibliography}

\end{document}